\documentclass[12pt,preprint]{aastex}

\usepackage{epstopdf}

\newcommand{\cmsq}{\hbox{cm$^{-2}$}}

\newcommand{\flux}{\hbox{erg~cm$^{-2}$~s$^{-1}$}}

\newcommand{\nh}{\hbox{${N}_{\rm H}$}}
\newcommand{\msun}{\hbox{${M}_{\odot}$}}

\newcommand\abs[1]{\left|#1\right|}

\newcommand{\be}{\begin{equation}}
\newcommand{\ee}{\end{equation}}
\newcommand{\ba}{\begin{eqnarray}}
\newcommand{\ea}{\end{eqnarray}}

\newcommand{\chandra}{{\emph{Chandra}}}

\newcommand{\xmm}{\emph{XMM-Newton}}

\newcommand{\rosat}{\emph{ROSAT}}

\newcommand{\swift}{\emph{Swift}}

\newcommand{\wise}{\emph{WISE}}

\newcommand{\simgt}{\lower 2pt \hbox{$\, \buildrel {\scriptstyle >}\over {\scriptstyle\sim}\,$}}
\newcommand{\simlt}{\lower 2pt \hbox{$\, \buildrel {\scriptstyle <}\over {\scriptstyle\sim}\,$}}
\newcommand{\ls}{\lower 2pt \hbox{$\;\scriptscriptstyle \buildrel<\over\sim\;$}}
\newcommand{\gs}{\lower 2pt \hbox{$\;\scriptscriptstyle \buildrel>\over\sim\;$}}
\newcommand{\sarc}{$^{\prime\prime}\!\!.$}
\newcommand{\marc}{$^{\prime}\!\!.$}

\begin{document}

\def\arcsec{$^{\prime\prime}$}
\def\arcmin{$^{\prime}$}
\def\degr{$^{\circ}$}

\title{The \emph{SWIFT} AGN and Cluster Survey I: Number Counts of AGN and Galaxy Clusters}

\author{Xinyu Dai\altaffilmark{1}, Rhiannon D. Griffin\altaffilmark{1}, Christopher S. Kochanek\altaffilmark{2}, Jenna M. Nugent\altaffilmark{1}, Joel N. Bregman\altaffilmark{3}} 

\altaffiltext{1}{Homer L. Dodge Department of Physics and Astronomy, University of Oklahoma, Norman, OK, 73019, xdai@ou.edu}
\altaffiltext{2}{Department of Astronomy and the Center for Cosmology and Astroparticle Physics, Ohio State University, Columbus, OH 43210}
\altaffiltext{3}{Department of Astronomy, University of Michigan, Ann Arbor, MI 48109}

\begin{abstract}
The \swift\ AGN and Cluster Survey (SACS) uses 125~deg$^2$ of \swift\ XRT serendipitous fields with variable depths surrounding $\gamma$-ray bursts to 
provide a medium depth ($4\times10^{-15}\flux$) and area survey filling the gap between deep, narrow \chandra/\xmm\ surveys and wide, shallow \rosat\ surveys.  
Here we present a catalog of 22,563 point sources and 442 extended sources and examine the number counts of the AGN and galaxy
cluster populations. 
SACS provides excellent constraints on the AGN number counts at the bright end with negligible uncertainties due to cosmic variance, and these constraints are consistent with previous measurements.
We use \wise\ mid-infrared (MIR) colors to classify the sources.  For AGN we can roughly separate the point sources into MIR-red and MIR-blue AGN, finding roughly
equal numbers of each type in the soft X-ray band (0.5--2~keV), but fewer MIR-blue sources in the hard X-ray band (2--8~keV). 
The cluster number counts, with 5\% uncertainties from cosmic variance, are also consistent with previous surveys but span a much larger 
continuous flux range.  Deep optical or IR follow-up observations of this cluster sample will significantly 
increase the number of higher redshift ($z > 0.5$) X-ray-selected clusters.
\end{abstract}


\section{Introduction}

X-ray (0.2--10 keV) surveys are fundamental tools to comprehensively study populations of X-ray sources
including active galactic nuclei (AGN), clusters and groups of galaxies, starburst galaxies, normal galaxies and their evolution.  
The most recent X-ray survey of the whole sky is the \rosat\ All Sky Survey (RASS, Voges et al.\ 1999) with a flux
limit of $3\times10^{-13}$\flux\ in the 0.1--2.4 keV band.  Surveys with \chandra\ and \xmm\ probe far
fainter fluxes over much smaller areas, and at this point more than 20 \chandra\ and \xmm\ surveys have been carried 
out with different combinations of survey area and depth ranging from the small \chandra\ deep fields
(Brandt et al.\ 2001; Xue et al.\ 2011) to the wider area XB\"ootes  (Murray et al.\ 2005) and XMM-LSS (Pierre et al.\ 2004)
surveys (see the review by Brandt \& Hasinger 2005, Figure~1).

Since even the largest \chandra\ and \xmm\ survey areas are $\sim 10$~deg$^2$, there is a need for intermediate depth,
wider area surveys to fill the gap between these small, deep surveys and the wider area, shallow surveys based 
on RASS.  Since (to zeroth order) survey volume scales as $V \propto \Omega f_{lim}^{-3/2}$ with survey area $\Omega$ and flux
limit $f_{lim}$, surveys of very different depths can have the same overall volumes.  They will differ, however,
in the redshifts they best probe.  RASS cluster surveys like the \rosat\ Brightest Cluster Sample (BCS, Ebeling et al.\ 2000) 
and the \rosat-ESO Flux Limited X-ray survey (REFLEX, Bohringer et al.\ 2004) are well-designed for studies
of clusters in the low redshift universe, while deep fields are well-designed for studies of lower luminosity, high redshift
AGN.  Shallow surveys, however, lack the depth to produce large samples at intermediate redshifts to probe
the evolution of clusters, and deep surveys tend to have their survey volumes at redshifts where clusters are
rare.  Similarly, shallow (deep) surveys characterize the bright (faint) AGN populations well, but both do
poorly for the intermediate luminosity populations and redshifts needed to link the two extremes.  Pencil
beam surveys, particularly at $z<1$, also are strongly affected by cosmic variance.  Filling this gap requires
surveying 100s of square degrees to intermediate depths.

Most existing medium-deep, wide area X-ray surveys are serendipitous surveys built from jointly analyzing
large numbers of archival pointed observations originally obtained for other purposes.  This includes 
\rosat\ pointed serendipitous surveys such as RIXOS (Castander et al.\ 1995), RDCS (Rosati et al.\ 1995; 1998), 
SHARC (Collins et al.\ 1997; Burke et al.\ 1997), WARPS (Scharf et al.\ 1997; Jones et al.\ 1998; Perlman et al.\ 2002), 
160 deg$^2$ (Vikhlinin et al.\ 1998), its extension 400 deg$^2$ (Burenin et al.\ 2007), ROXS (Donahue et al.\ 2001), and 
BMW (Campana et al.\ 1999).  Similarly, the ChaMP Survey (Kim et al.\ 2004) pursues the same goals using
archival \chandra\ data.  There are exceptions, like the RASS NEP (Henry et al.\ 2001) survey which was based
on the repeatedly scanned North Ecliptic Pole regions.
The \rosat\ surveys still provide the best constraints on the bright end of the cluster X-ray luminosity function and the high redshift cluster mass function
(Rosati et al.\ 2002; Vikhlinin et al.\ 2009) due to their large survey area.
The \xmm\ and \chandra\ serendipitous surveys are still on going (e.g., Watson et al.\ 2009; Lloyd-Davies et al.\ 2011; Fassbender et al.\ 2011; Clerc et al.\ 2012).
These surveys cover sky areas of 10--400 deg$^2$, and have flux limits from $5\times10^{-14}$ to $10^{-15}$~\flux.
One disadvantage of most of these serendipitous surveys based on pointed archival data is that the selection effects 
are difficult to model without a complete understanding of why the original targets were selected.
Figure~1 shows where some of these surveys lie in the space of area and depth.

\begin{figure}
\plotone{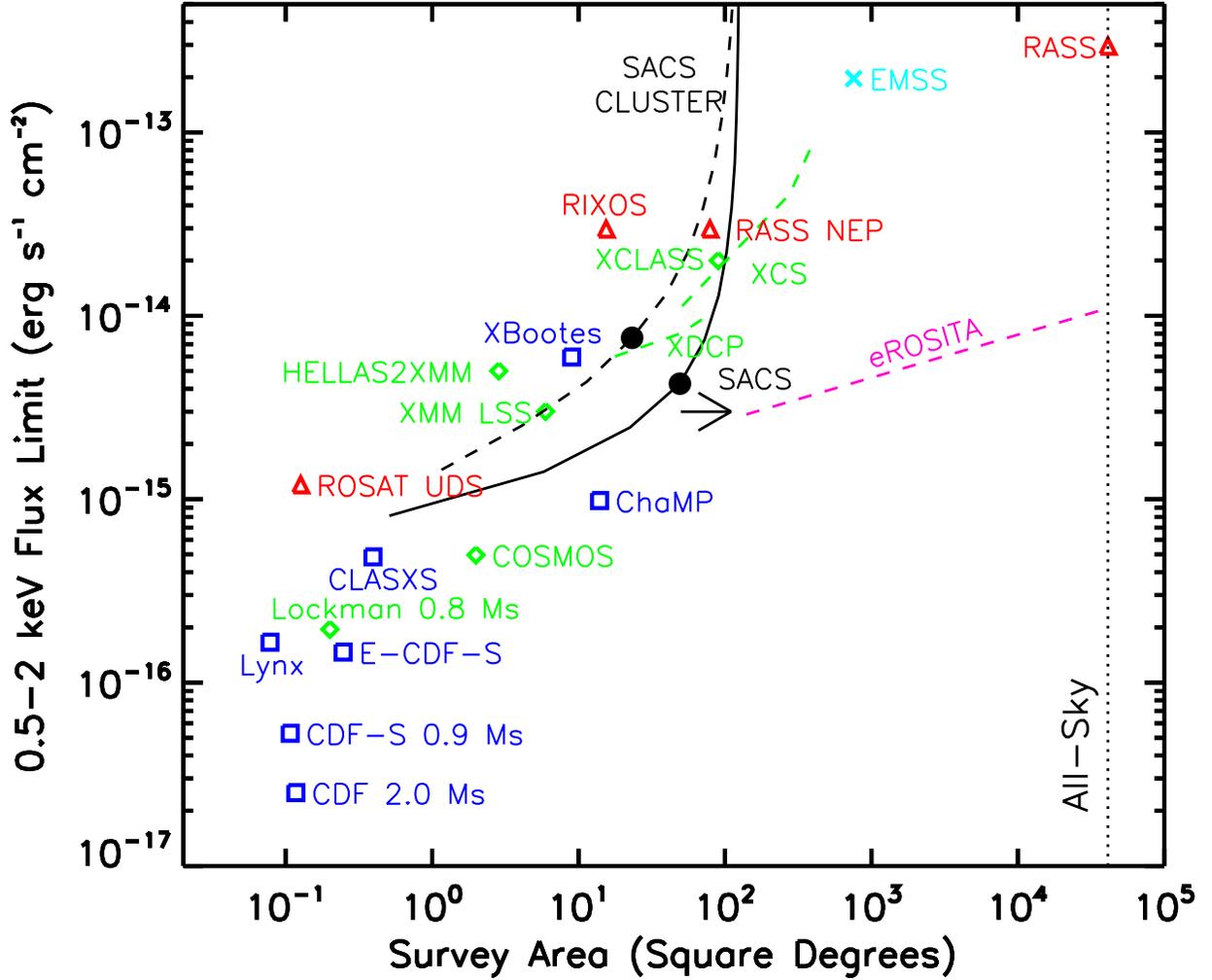}
\vspace{1 cm}
\caption{
  Comparison of the flux limit and survey area of the \swift\ AGN and Cluster Survey (SACS) through July 2013 to
  other soft X-ray surveys from Brandt \& Hasinger (2005), adding recent \xmm\ medium deep surveys, and the 
  future \emph{eROSITA} surveys.  SACS (solid and dashed lines for point and extended sources) is substantially wider or deeper than 
  many similar existing surveys, and the arrow shows that SACS is approaching the \emph{eROSITA} deep survey as \swift\ continues to accumulate data.
The solid circles indicate the typical SACS field, 49 square degrees and $4.3\times10^{-15}~\flux$ for point sources and 23 square degrees and $7.6\times10^{-15}~\flux$ for extended sources, in the 0.2--5~keV band.
\label{fig:fluxarea}}
\end{figure}

\begin{figure}
\plotone{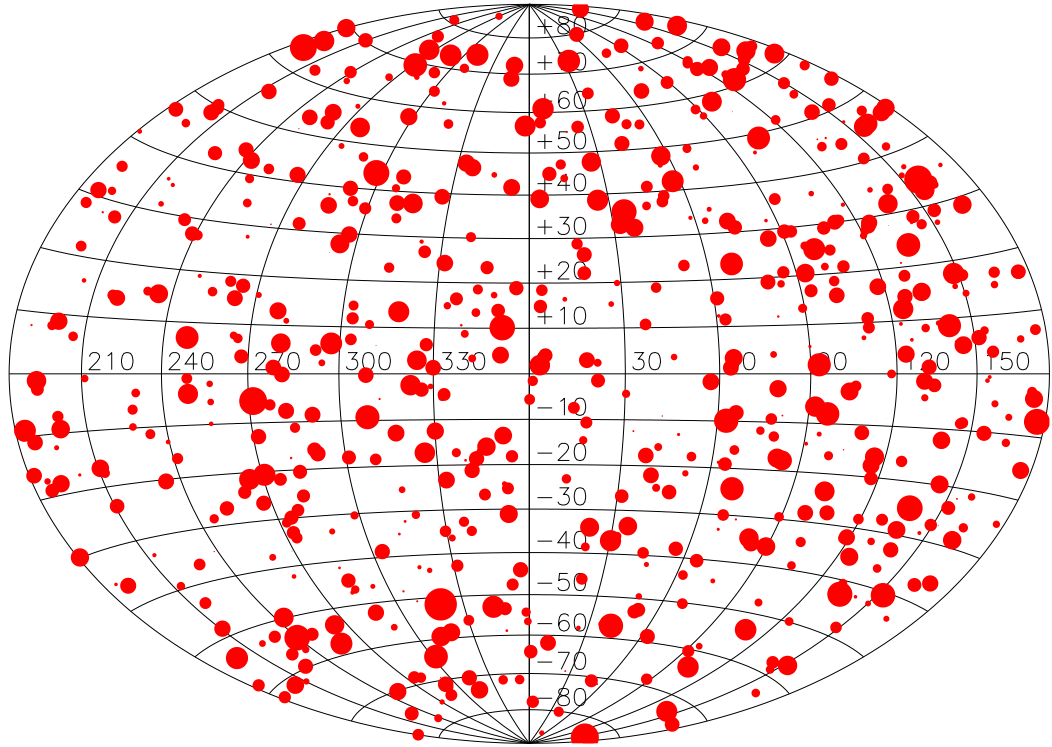}
\caption{Distribution of \swift--XRT GRB observations in Galactic coordinates.  Larger point sizes indicate longer exposure times. \label{fig:sky}}
\end{figure}

In this paper, we present a serendipitous medium-deep, wide-field soft X-ray survey using \swift\ (Gehrels et al.\ 2004; 
Burrows et al.\ 2005; Morette et al.\ 2005) observations of $\gamma$-ray bursts (GRBs).  
From its launch in November 2004 through July 2013, \swift\ has observed 784 GRBs at a rate of $\sim92$ bursts 
per year.  The X-Ray Telescope (XRT) on board \swift\ is sensitive in the energy range from 0.2--10 keV and has 
a field of view of 23.4$\times$23.4 arcmin$^2$.  These GRB fields are randomly distributed on the sky (Figure~\ref{fig:sky}) 
and are essentially uncorrelated with other X-ray source populations.  Thus, the 
XRT observations of GRB fields form an excellent soft X-ray serendipitous survey covering a total sky area of 
$\sim 125$ deg$^2$ with variable depths and a median flux limit of $4\times10^{-15}$ \flux.
Figure~\ref{fig:fluxarea} shows the relative depth and sky coverage of these \swift\ serendipitous fields 
compared to other surveys.  This unique combination of the survey area, survey depth and ``randomness''
enables it to fill in the gap between deep, pencil beam surveys such as the \chandra\ Deep Fields 
and the shallow, wide area \rosat\ surveys.  Thus it can be used to make independent and complementary measurements 
of the number counts and luminosity functions of X-ray sources, principally AGN and galaxy clusters.
This will in turn strengthen our understanding of the evolution of X-ray sources and the underlying cosmology.
Several other groups are also independently working with this data, focusing on bright clusters 
(Tundo et al.\ 2012; Liu et al.\ 2013), point sources in deep XRT fields (Puccetti et al.\ 2011), and on
the overall source catalog of all XRT observations (D'Elia et al.\ 2013; Evans et al.\ 2014).  

In \S2 we describe our field selection, the reduction of the XRT data and the detection of sources. In \S3
we classify the sources based on the X-ray angular extent and matches to the Wide-field Infrared Survey Explorer
(\wise, \citealt{Wright2010}) survey catalogs. In \S4 we compare our catalogs with other independent \swift\ XRT catalogs.  
In \S5 we compare the source number counts to existing estimates
from both shallower and deeper surveys, and we discuss the results in \S6.  We assume cosmological parameters of 
$\Omega_M = 0.27$, $\Omega_{\Lambda}=0.73$, $\sigma_8 = 0.81$, and $H_0 = 70$~km~s$^{-1}$~Mpc$^{-1}$ throughout 
the paper.

\begin{figure}
\epsscale{0.8}
\plotone{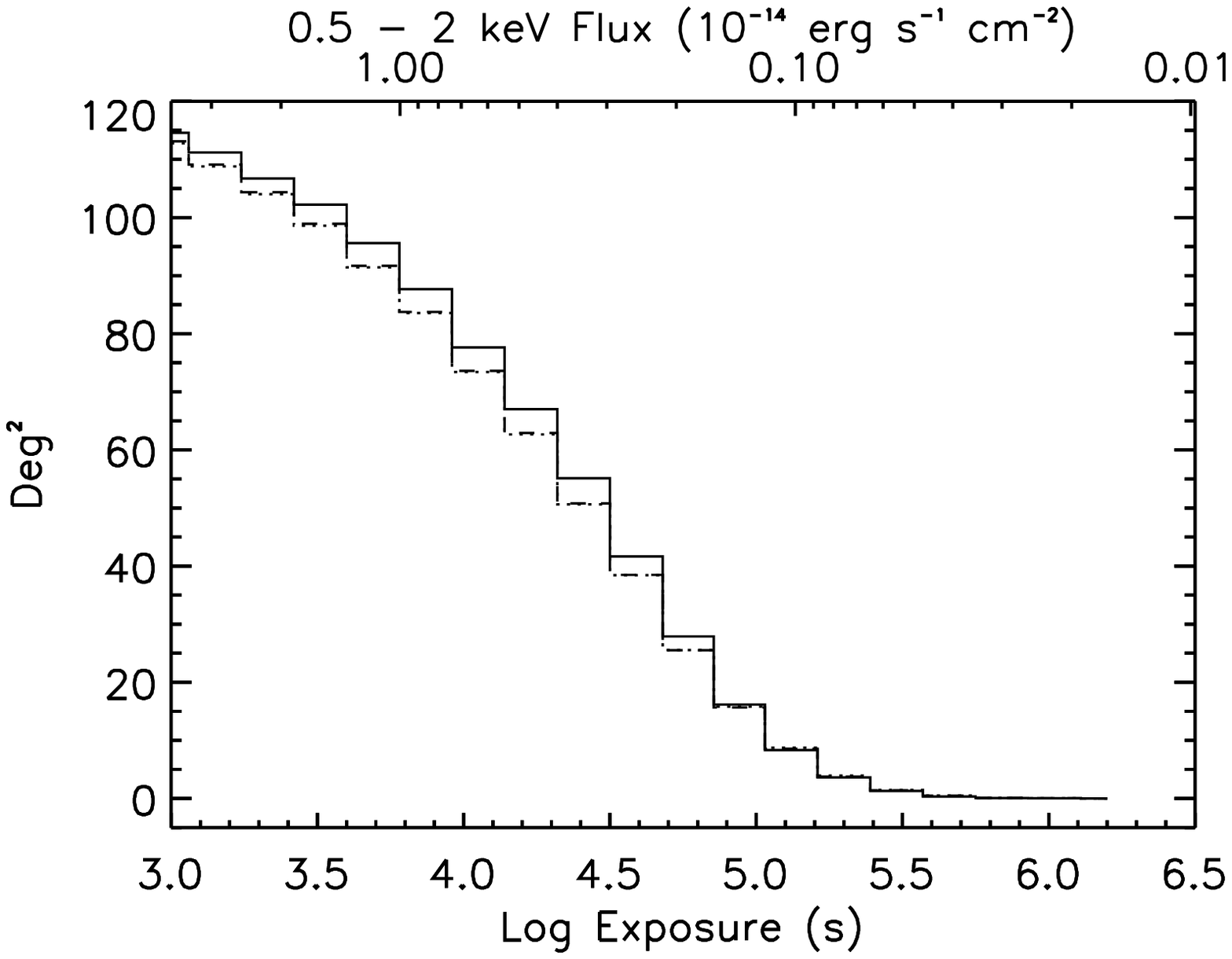}
\vspace{0.0cm}
\plotone{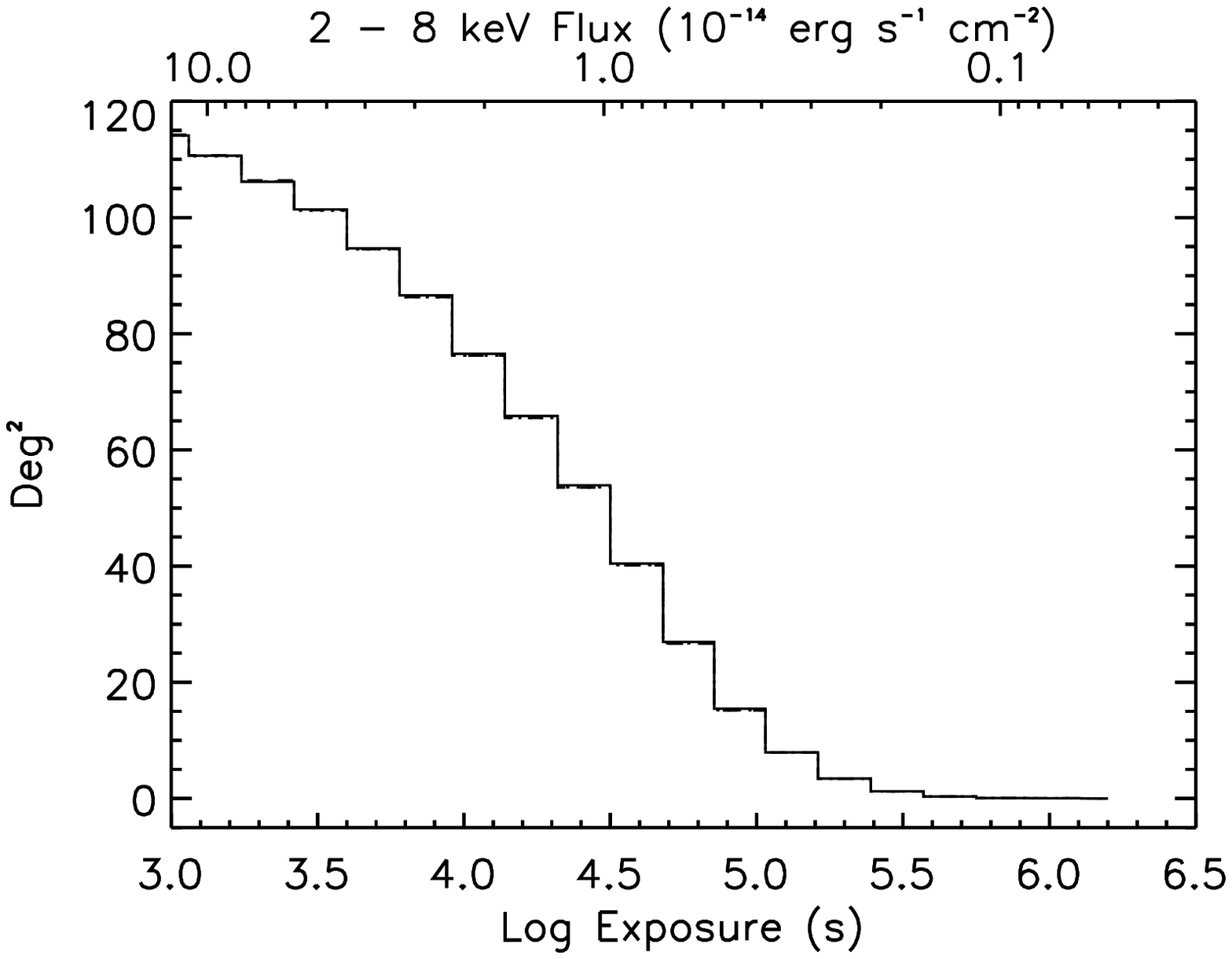}
\vspace{-0.5cm}
\caption{ 
   Cumulative distribution of survey area as a function of the equivalent on-axis exposure time 
  for the soft (0.5--2~keV, upper panel) and hard (2--10~keV, bottom panel) X-ray bands.  The dotted 
  and dashed lines show the effect of smoothing the exposure maps over 6 and 10 XRT pixels, the typical
  sizes of point and extended sources, and with a nominal Galactic \nh\ of $5\times10^{20}\cmsq$.  
  In the hard band, the curves are almost completely overlapping.
 The upper X-axis shows the corresponding flux limits based on the median background rates of the fields and a typical Galactic \nh\ of $5\times10^{20}\cmsq$.
  \label{fig:areaint}}
\end{figure}

\begin{figure}
\epsscale{1}
\plotone{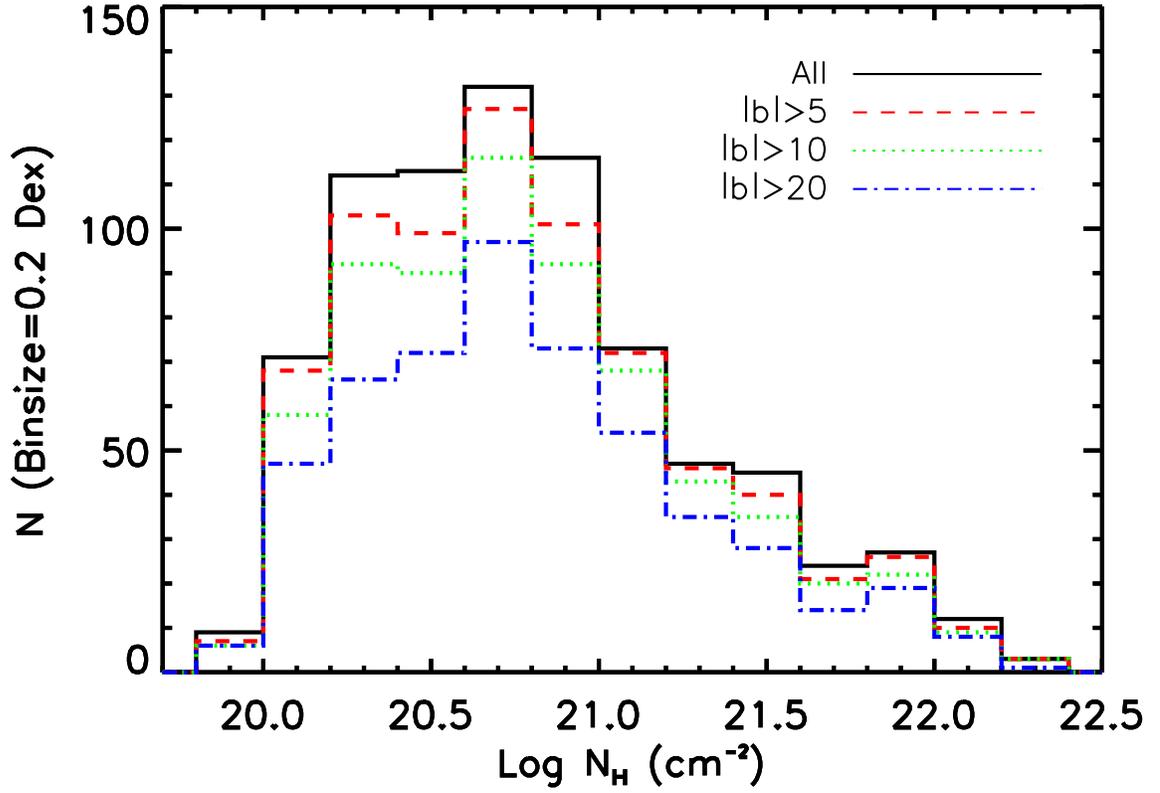}
\caption{ 
   Galactic \nh\ column density for all of the \swift\ fields (black solid line), 
   and those with Galactic latitudes of $\abs{b} \ge 5$ (red dashed line), $\abs{b} \ge 10$ (green dotted line) , and $\abs{b} \ge 20$ (blue dot-dashed line),
   respectively. \label{fig:nh}}
\end{figure}

\section{Data Reduction and Source Detections}

The \swift\ XRT observations of GRB fields were downloaded from the HEASARC website\footnote{The HEASARC website is at http://heasarc.gsfc.nasa.gov.}.  
This includes all XRT observations before 2013--07--27 with target names containing ``GRB'', but excluding special targets with names such as 
``Non-GRB''.  We lose the very small fraction ($<1\%$) of XRT GRB observations without ``GRB'' in their target names
(usually instead having target names starting with ``Swift~J'').  We reprocessed the data using the HEASoft tool \verb+xrtpipeline+ 
(version 0.12.6) and \swift\ CALDB version XRT20130313.  We generated the corresponding exposure maps using \verb+xrtexpomap+ (version 0.2.7).  
XRT follow-up observations of GRBs are typically a monitoring sequence to measure the decay of the X-ray light emission, so we
merged all the reprocessed event files and exposure maps for each GRB field into a single event file and a single exposure map in
order to increase the signal-to-noise ($S/N$) ratio for detecting serendipitous sources.
We then created images in the total (0.2--10~keV), soft (0.5--2~keV), and hard (2--10~keV) bands from the merged event files.
Figure~\ref{fig:areaint}
shows the cumulative area of the final field sample as a function of the equivalent on-axis exposure time and Figure~\ref{fig:nh} shows 
the distribution of the fields in Galactic \nh\ column density based on \citet{dl90}.  
Since the \nh\ distribution for XRT fields with low Galactic latitudes, $\abs{b} < 5^\circ$, is similar to those from high Galactic latitudes, we included all these XRT fields in our analysis. 

We used a wavelet-based algorithm (\verb+wavdetect+, Freeman et al.\ 2002) in CIAO to identify sources, an algorithm which is widely used in
the X-ray community to detect sources in \chandra\ and \rosat\ images.   Since one of our goals is to separate 
point AGN from extended clusters, the angular structure of the \swift/XRT point spread function (PSF) has several advantages.
First, its dependence on off-axis angle and energy are well measured by the \swift\ team \citep{moretti05} and released as a part 
of the calibration files.  Second, the on-axis half power diameter (HPD) of 17--18\arcsec\ is reasonably well-suited for cluster
identification. More importantly it is roughly constant over the full field of view, unlike \chandra\ and \rosat \citep{weisskopf02, hasinger94} where they are very strong
functions of off-axis angles.  For example, at 1.5 keV, the on-axis HPD is 18\sarc3, slowly declining to 15\sarc1 for a 10\arcmin\ off-axis angle. 
For \chandra\ and \rosat, the HPDs of 12\sarc4 and 21\arcsec at 10\arcmin\ off-axis are much larger than the on-axis HPDs of 0\sarc84 and 14\arcsec.
The PSF of \xmm\ has smaller differences, but still increases from 15\arcsec\ (13\arcsec) on-axis to 19\arcsec\ (21\arcsec) at 10\arcmin\ off-axis for the PN (MOS) cameras \citep{jansen01}.  
The relatively uniform PSF also significantly reduces systematic uncertainties when assessing the significance of 
source detections and their angular sizes. 
We used \verb+wavdetect+ to detect sources in the total, soft, and hard images independently 
using a false positive threshold of $10^{-6}$.  
We further evaluated the significance of the detections using the net source and background counts provided by \verb+wavdetect+ and Equation A11 of \citet{weisskopf07} for Poisson statistics,
\begin{equation}
P = \Gamma(N_{Total}, N_{Bkg})/\Gamma(N_{Total}), 
\end{equation}
where $N_{Total} =  N_{Src,\ Net} + N_{Bkg}$, and rejected sources with significance less than 0.997 from the final catalog. 
These final catalogs are presented in Tables~1--4, where we have used the results of the next section
to separate them into candidate AGN and galaxy clusters.
Tables~1 to 3 present the 22,563, 17,748, and 10,060 total, soft, and hard band AGN candidates and Table~4 presents the 442 cluster candidates.
We corrected the fluxes for the finite aperture used in the measurements. 
For point sources we simply use the PSF, while for the extended sources we assumed a $\beta=0.6$,
$R_c=0.1$~Mpc $\beta$ model \citep[e.g.,][]{sanderson03} at the typical cluster redshift of $z=0.5$ 
and estimated the total flux inside 2~Mpc.  This typically increased the fluxes by a factor
of $1.3$.  Similarly, we assume a $T=5$~keV, $Z=0.4Z_{\odot}$ plasma spectrum at $z=0.5$ 
to correct their fluxes for Galactic absorption.  
In the number count calculations (Section~\ref{sec:logns}), we set the Galactic absorption as $\nh = 5\times10^{20}\cmsq$ for all sources; 
however, we adopted the \citet{dl90} Galactic absorption values when reporting the fluxes in Tables~1--4.

\section{Classification of the X-ray Sources}

The vast majority of our sources will be extragalactic sources, AGN, GRBs and galaxy clusters, since only a small fraction of
XRT observations are in the Galactic plane and Galactic X-ray sources are typically fainter.
Our matching analysis with \wise\ sources also suggests that there are few Galactic sources in our catalogs (Section~\ref{sec:wise}).
In this section we discuss the classification of the sources using their X-ray angular extent and their
properties, where detected, in the all-sky \wise\ catalogs.  We first matched our sources to the known
positions of the GRB X-ray afterglows using the most recent catalogs from the UK XRT 
team\footnote{Available at http://www.swift.ac.uk/xrt\_positions/.}.  We used a matching radius of 24\arcsec\
even though the \swift\ XRT HPD is about 18\arcsec.  The XRT position of GRBs can be measured much more 
accurately because of the large number of photons detected in GRBs.  However, for bright
GRBs a large fraction of photons located near the center of the GRB image can be rejected during the
data reduction process because of strong pile-up effects.  These sources are then found by \verb+wavdetect+
as a source pair slightly offset from the true position but always within 24\arcsec\ of each other.  
Since the GRB science is not the main focus of this paper and to avoid its strong contamination to other sources, 
we remove all sources within 24\arcsec\ of the GRB positions. 
The next step is to use the angular source extent to distinguish between AGN and galaxy clusters.

\subsection{Extended Source Analysis\label{sec:ext}}

All galaxy
clusters are extended X-ray sources given an ideal telescope.  Cluster surface brightness profiles are well-modeled
by the so-called $\beta$ profile, $S \propto  \left(1+\left(R/R_c\right)^2\right)^{-3\beta+1/2}$ \citep[e.g.,][]{xu01, sanderson03, dai07, dai10}.    
The core radius $R_c$ and power law index $\beta$ depend on cluster
mass \citep{xu01, sanderson03, dai07, dai10}, and values of (0.2~Mpc, 0.67), (0.1~Mpc, 0.6) and (0.05~Mpc, 0.5) are typical for clusters
with masses of $10^{15}$, $2\times10^{14}$ and $5\times10^{13}M_\odot$ \citep[e.g.,][]{na99, xu01, sanderson03, op04, dai10}.  The \swift\ XRT PSF has a moderate
HPD ($15$--$18$\arcsec) and is also well-modeled by a $\beta$ profile with $\beta \simeq 0.67$
and $R_c \simeq 2.3$ XRT pixels with variations across the field of view that are well understood (Moretti et al.\ 2005).\footnote{
In Moretti et al. (2005), this is described as being a King profile, $\left(1+\left(R/R_c\right)^2\right)^{-\alpha}$, which is
just a $\beta$ model with $\alpha=3\beta-1/2$.}  The observed profile of a cluster is then the
convolution of the appropriate $\beta$ model with the PSF.  For simplicity, we will simply view
this as a $\beta$ model with different parameters.  

We used only the soft band images to study the source surface brightness profiles because of their higher contrasts for cluster emission, and used an iterative procedure in this analysis.
From the merged soft X-ray image of each field, we extracted background-subtracted surface brightness profiles of each source 
 out to the radius $r_b$, 
where the surface brightness equals half the background level and the initial guesses were set to two times the source radii given by \verb+wavdetect+.  
 We then fit these surface brightness profiles with $\beta$ models, with $\beta =0.67$ fixed for all sources.
This is optimal
for the vast majority of point sources and massive clusters.
Sources were analyzed in the order of their brightness, and when we analyzed 
sources we included the flux contributions from other nearby sources.  We started with the brightest
source, fixing the parameters of all other sources in the model when fitting each individual source, and sequentially moved down to the faintest source.  
We then iterated this procedure until the model parameters ($R_c$ and total flux) for all the sources converge.

We analyzed simulated sources to estimate the detection thresholds for extended sources by adding
1000 simulated clusters to real images and then using our detection algorithm to recover
the extended sources and their properties.  
Similarly, we added the same number of simulated point sources
and recovered their properties.  
Here, we used the $R_c$ and $\beta$ pair of (0.1~Mpc, 0.6) for clusters with total masses of $\sim 2\times10^{14}~\msun$,
which correspond to our typical cluster detection limit.
Figure~\ref{fig:sig} shows an example of the probability of 
distinguishing extended and point sources as a function of the net number of photons for images
with a typical exposure time of 60~ks.   For the significance of the difference to exceed
3$\sigma$, it requires 17--30 net photons for clusters from $z=0.6$--$1.4$, which 
correspond to fluxes of 0.6--$1.1\times10^{-14}$ \flux\ in the 0.5--2 keV band for a typical XRT field.  
Higher numbers of photons are needed for higher redshift clusters
as the angular sizes of the clusters become smaller.  
More massive clusters have larger core sizes (e.g., $R_c = 0.2$~Mpc), which are easier to separate from point sources.
Using the same $R_c$ and $\beta$ pair (0.1~Mpc, 0.6), we carried out these simulations for
the range of backgrounds encompassing the shallowest and deepest observations and these can
be used to model the completeness of the cluster sample (Table~\ref{tab:clprob}).  

Figures~\ref{fig:off} and \ref{fig:sn} show the distribution of sizes ($R_c$) for the sources
detected in the XRT images as a function of the off-axis angle and the signal-to-noise ratio
of the detection.  There is a dominant band of points sources with $R_c \simeq 2.5$~XRT pixels (6\arcsec), 
as expected from the PSF models.  The mean size of the point sources is essentially independent of the
off axis angle, as expected from the uniformity of the PSF, and the scatter in the sizes increases
for lower signal-to-noise ratios.  To estimate the boundary for selecting extended sources we set
an initial boundary that roughly separated the point and extended sources and then calculated
the mean and standard deviation of the point source sizes.  We then updated the $3\sigma$ boundary and
iterated this procedure until the estimates converged.  This was done as a function of both
off-axis angle and signal-to-noise ratio with the results also shown in Figures~\ref{fig:off} 
and \ref{fig:sn}.  We defined the extended source catalog as sources with $S/N \geq 4$ and a minimum net photon count of 20 that
are more than 3$\sigma$ from the mean size of point sources for both their off-axis angle 
and signal-to-noise ratio.   This resulted in 442 cluster candidates.  

We matched our cluster catalog with the meta X-ray cluster catalog (Piffaretti et al.\ 2011) 
using a match distance of 0\marc5, and found seven matches.  In particular, SWCL J025630.7+000601, 
J062915.2+460619, J155743.3+353020, J180228.7-523651, J181628.8+691131, J214359.4-563725, and J233616.8-313629
match with MCXC J0256.5+0006, MCXC J0629.1+4606, MCXC J1557.7+3530, MCXC J1802.4-5236, MCXC J1816.5+6911, 
MCXC J2143.9-5637, and MCXC J2336.2-3136, respectively.
We also matched to the ACT and {\emph{Planck}} SZ cluster catalogs \citep{marriage11, planck14} and found that SWCL 025630.7+000601 matches ACTCL J0256.5+0006 
within 0\marc5 and SWCL J084749.4+133142 matches with PSZ1 G213.43+31.78 within 2\marc9. 
We also found 22 matches to the optical cluster catalogs \citep{koester07, hao10} in the SDSS footprint, and the details are
discussed in Griffin et al.\ (2015, in preparation).
We matched with the independent \swift\ cluster catalog of \citet{tundo12} with 72 entries and found 55 matches (see Section~\ref{sec:com}).
Therefore, the majority of our \swift\ clusters are new discoveries.

\begin{figure}
\plotone{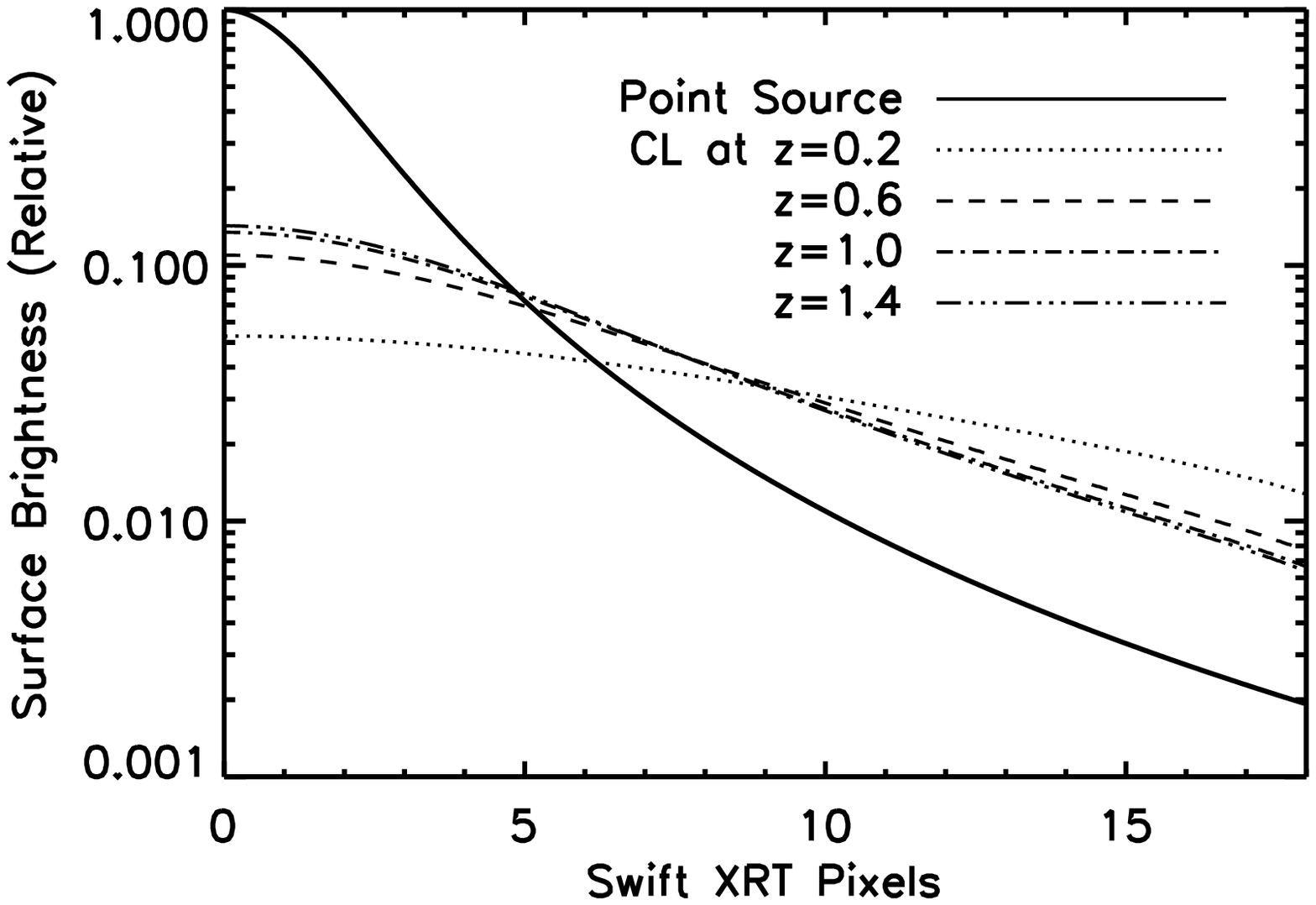}
\caption{Simulated surface brightness profiles for a point source (solid line) and a cluster with the same flux ($R_c=100$~kpc, $\beta=0.6$) 
 at redshifts of $z=0.2$, $0.6$, $1.0$ and $1.4$ convolved with the \swift--XRT PSF.  Even at redshift
 $z=1.4$, clusters are clearly more extended than point sources.
  \label{fig:psf}}
\end{figure}

\begin{figure}
\plotone{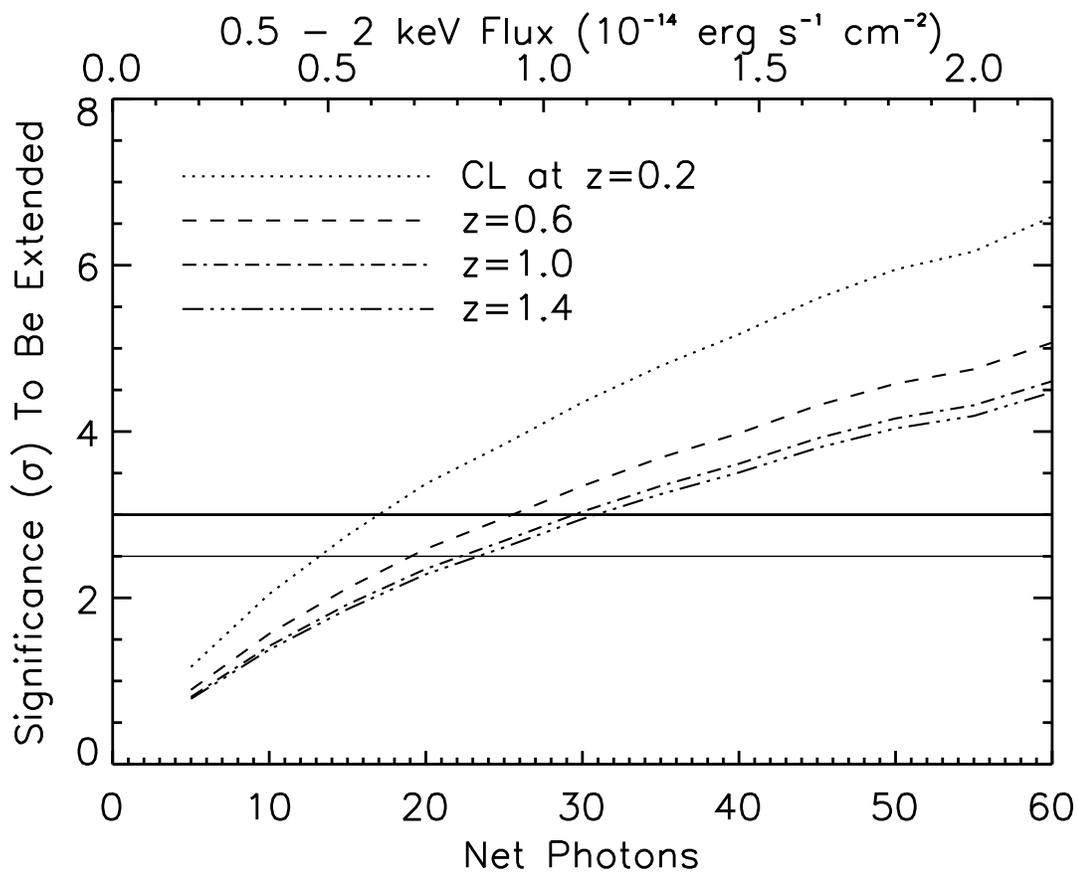}
\caption{Significance with which a cluster ($R_c=100$~kpc, $\beta=0.6$) can be detected for
  redshifts of $z=0.2$, $0.6$, $1.0$ and $1.4$ as a function of the net number of photons
  in an image with a typical exposure time of 60~ks.  The two horizontal solid lines 
  show the 2.5$\sigma$ and 3$\sigma$ significance levels. 
  The upper X-axis shows the corresponding flux. \label{fig:sig}}
\end{figure}

\begin{figure}
\plotone{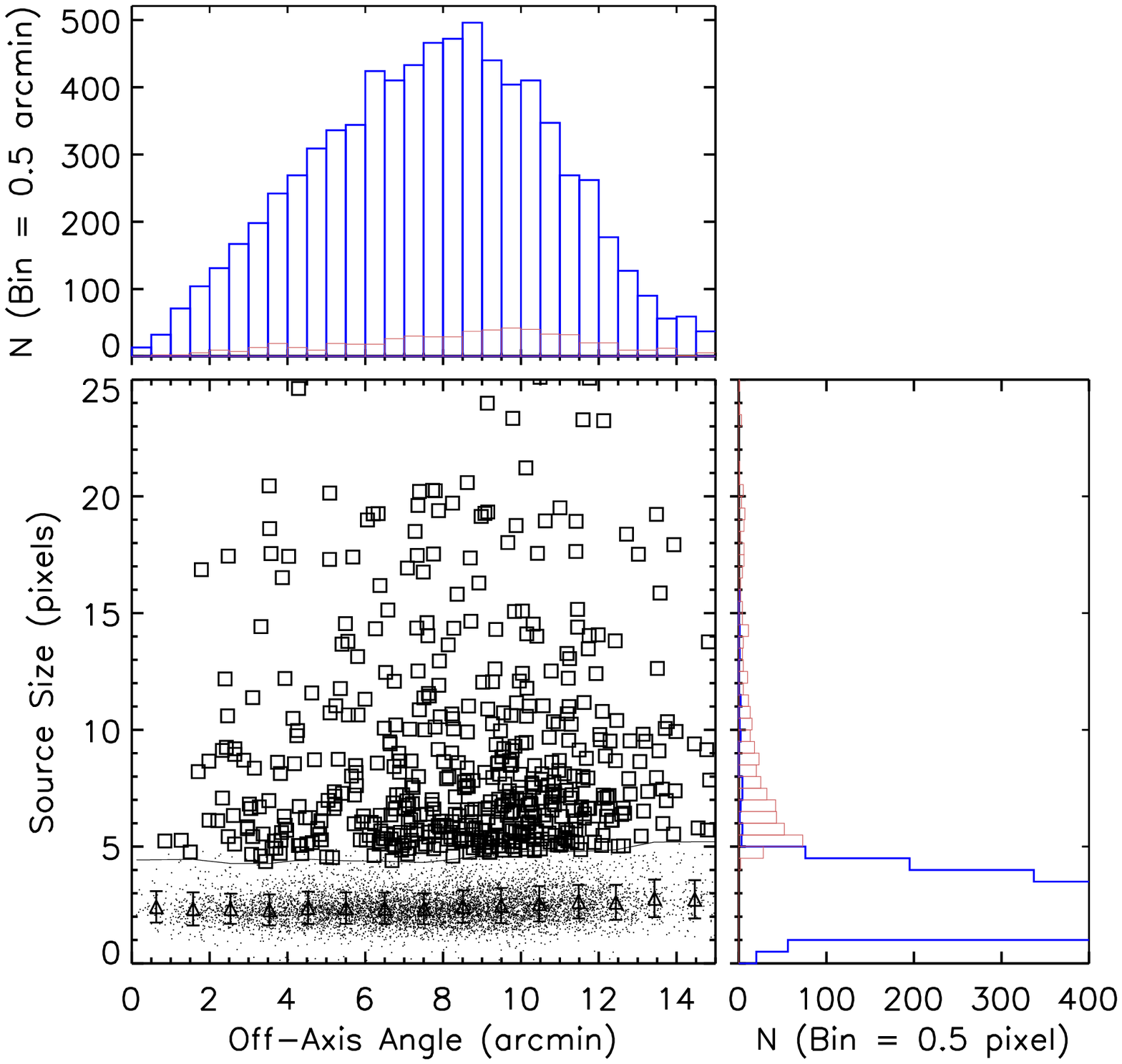}
\caption{The core radius $R_c$ distribution for the 17 thousand 0.5--2~keV sources detected in the 
  \swift\ GRB fields as a function of the source off-axis angle.  The triangles with
  error bars show the mean size of the point sources and the standard deviation about
  the mean.  The line shows the $3\sigma$ size cut used to separate point sources
  (solid line) from extended sources (open squares). 
The upper and right histograms show the $R_c$ and off-axis angle histograms for point sources (blue) and extended sources (red).
\label{fig:off}}
\end{figure}

\begin{figure}
\plotone{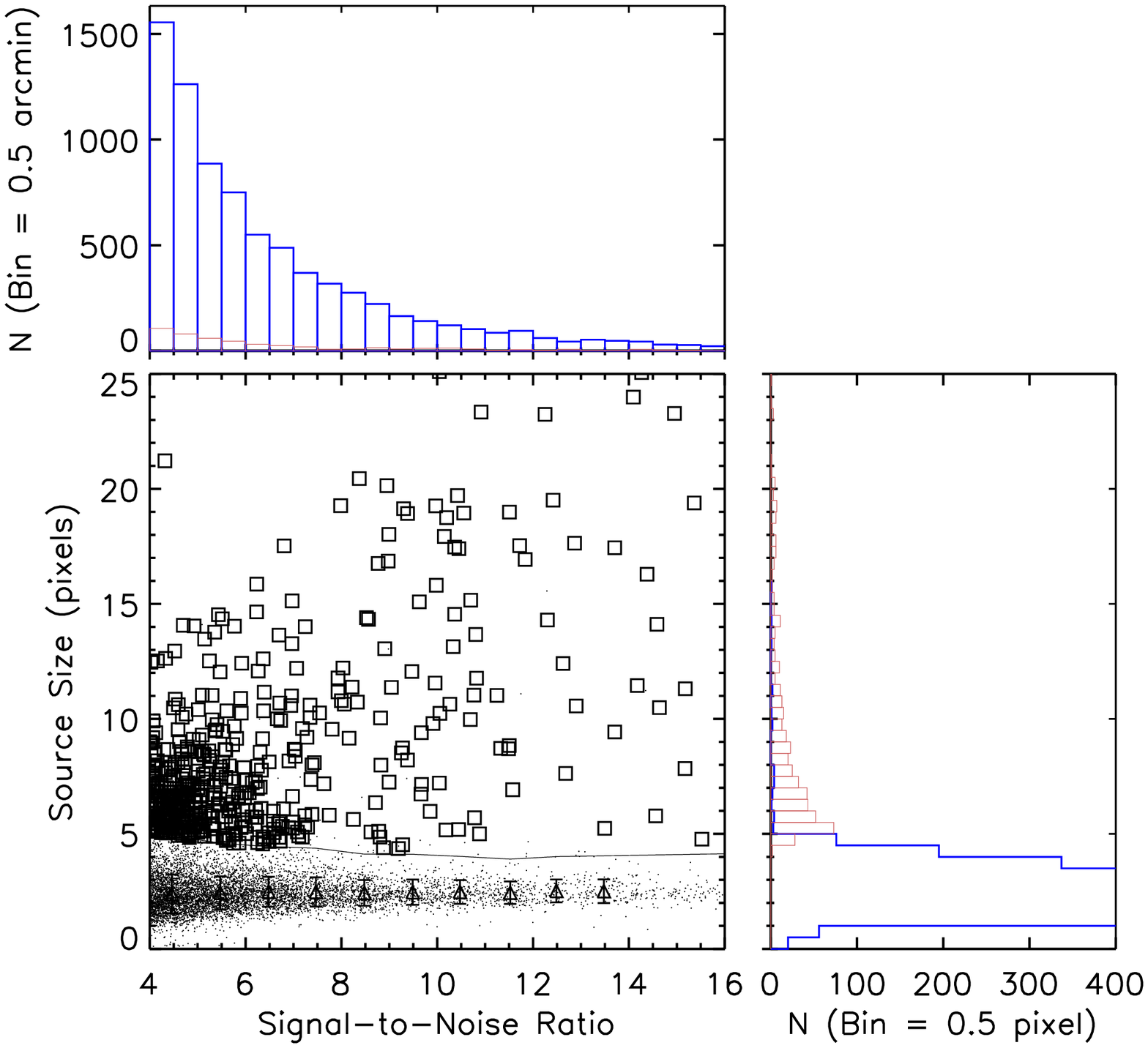}
\caption{The core radius $R_c$ distribution of the 17 thousand sources detected in the 
   \swift\ GRB fields as a function of the S/N.  As in Figure~\ref{fig:off} the 
   triangles with error bars show the mean size of the point sources and the standard deviation about
  the mean.  The line shows the $3\sigma$ size cut used to separate point sources
  (solid line) from extended sources (open squares).
The upper and right histograms show the S/N and $R_c$ histograms for point sources (blue) and extended sources (red).
\label{fig:sn}}
\end{figure}

\subsection{Matching to \wise\ \label{sec:wise}}

The only all-sky catalog with a depth well-matched to our survey is the mid-infrared (MIR) \wise\ \citep{Wright2010}
survey, which also automatically includes matches to the shallower \emph{2MASS} survey \citep{Skrutskie2006}.  
There are significant overlaps with other surveys, particularly SDSS, but nothing else provides
uniform coverage.   In Griffin et al.\ (2015) we examine the properties of the lower redshift
($z<0.5$) cluster candidates in the SDSS regions, but deeper data than SDSS are needed study
cluster properties in general.  Mid-IR colors are also a powerful and well-understood means of identifying
AGN (e.g., \citealt{Stern2005}) including examinations of AGN selection from \wise\ data
(e.g. \citealt{assef13}).  X-ray emission and MIR colors do not select identical
AGN samples.  Like X-rays, MIR selection is fairly insensitive to absorption.  However,
it does depend on the slope of the MIR spectrum being different from that of a galaxy,
which means that the AGN must make a significant contribution to the total MIR luminosity.
As a result, MIR selection will not identify very low luminosity AGN.  \cite{Chung2014}
has an extensive comparison of X-ray and MIR selection methods.  
We also improve the astrometry of the \swift-XRT sources by correcting the pointing accuracy of
XRT (3\arcsec, Burrows et al.\ 2005) during the \wise\ matching analysis.

\begin{figure}
\epsscale{0.8}
\plotone{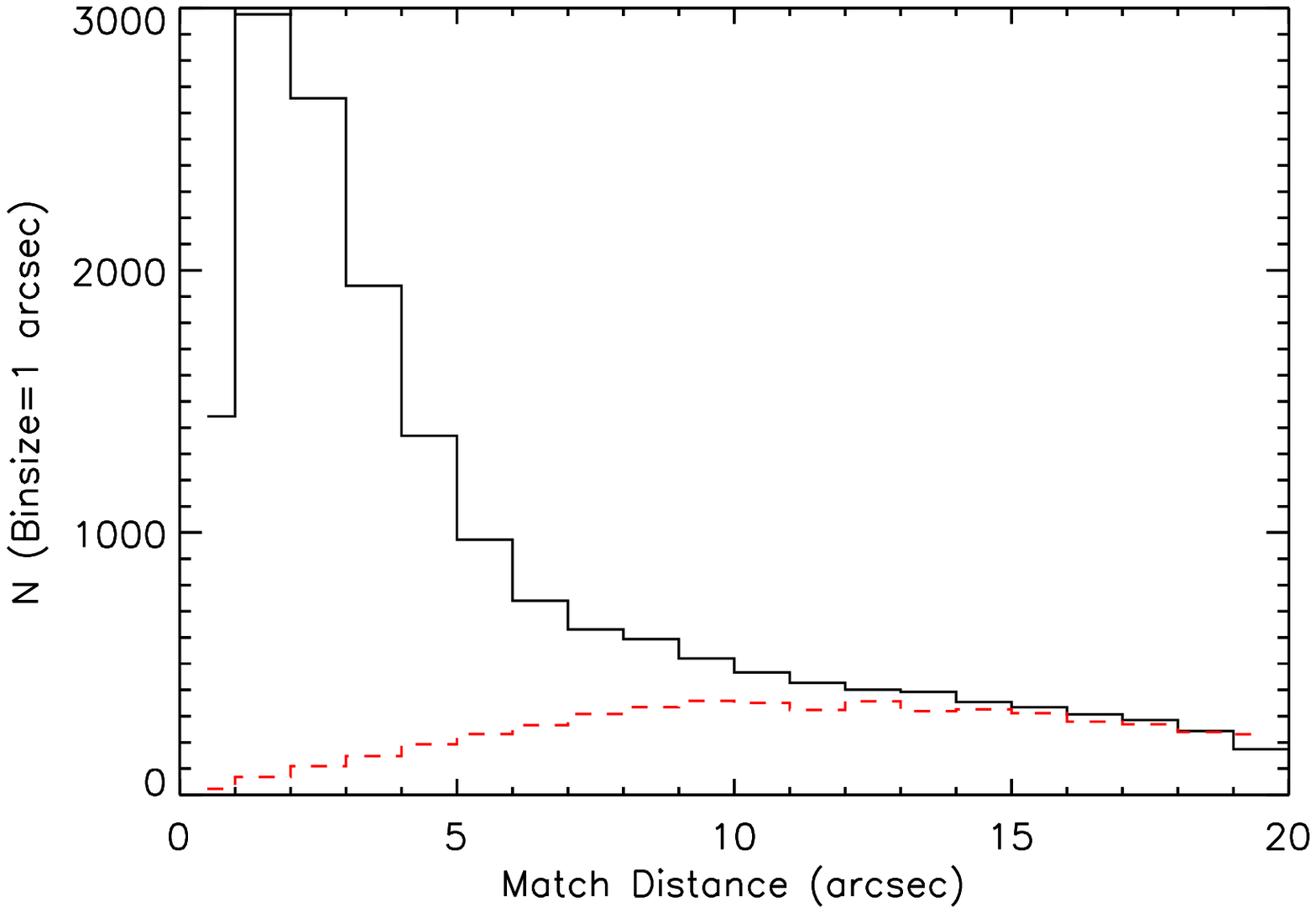}
\plotone{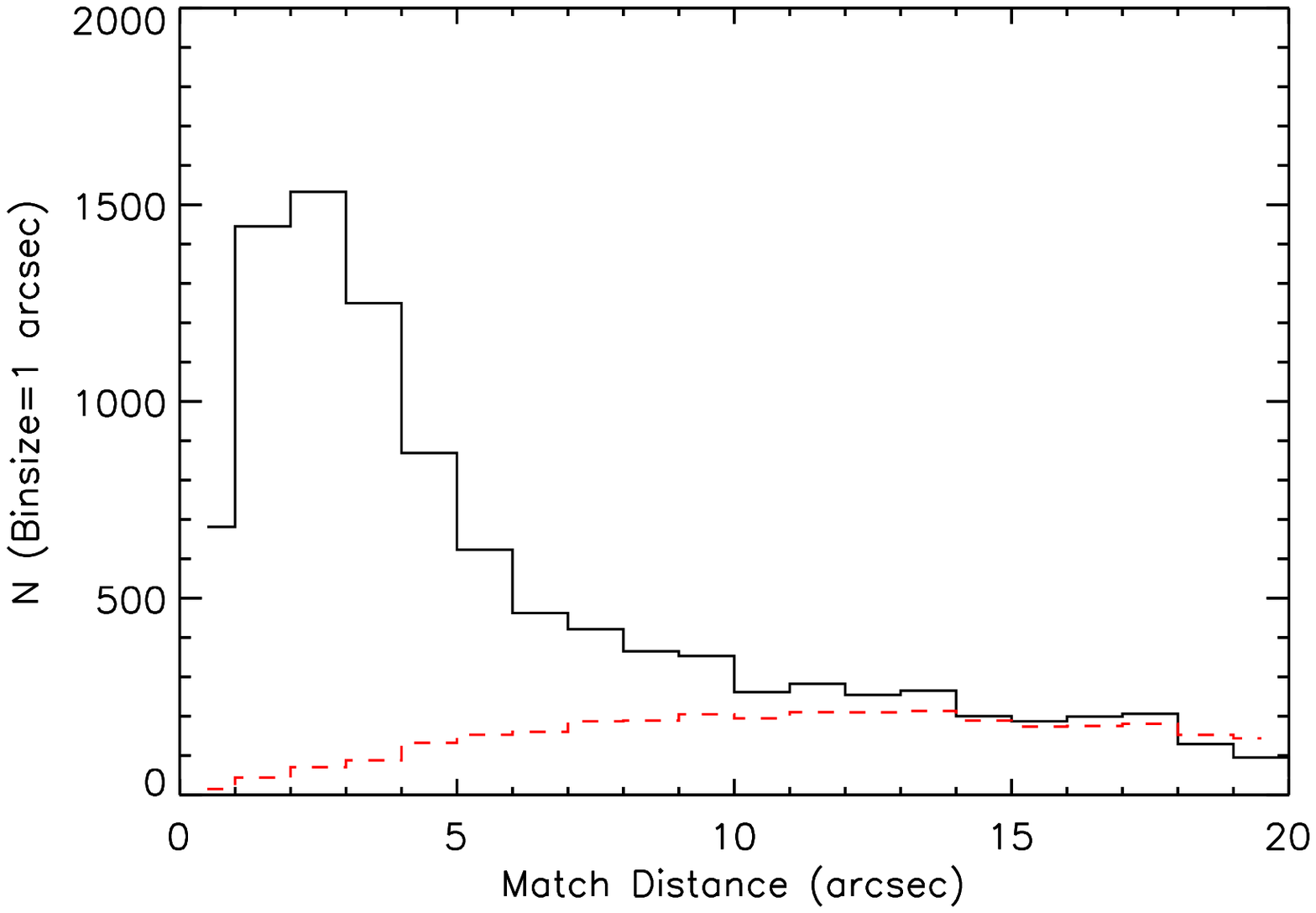}
\caption{Histograms of the distance between the soft (top) and hard (bottom) band \swift\ X-ray source and the closest \wise\ source. 
The black solid histogram is for the real X-ray sources, while the red dashed line is for nearby random positions.
   \label{fig:mdsoft}}
\end{figure}

We searched for the closest source to the position of each XRT source and to random
positions along a circle of radius 120\arcsec\ around the XRT position.  We kept the closest
\wise\ source to both the real and random positions provided the separation was
smaller than 20\arcsec.  
For each GRB field, we calculated the mean $\Delta$RA and $\Delta$Dec and variance of the these means for real matches, 
and corrected the X-ray source astrometry if any mean offset is larger than the standard deviation of the mean offsets.
The median and maximum RA corrections are 1\sarc36 and 5\sarc01, and the median and maximum Dec corrections are 1\sarc31 and 4\sarc36.
The top and bottom panels of Figure~\ref{fig:mdsoft} show the
distributions of the match distances for real and random sources after correcting the XRT astrometry and after normalizing
the random distribution to match the real distribution on scales of $15$--$20$\arcsec.
For the real sources we see a central core peaking at $\sim 2$\arcsec, with a non-zero peak, as expected from a $\chi^2$ distribution.
The probability that a match is correct as a function of distance
can be roughly estimated as $1-N_{fake}/N_{real}$ and we also show this probability
as a function of match distance.  
The match probabilities are roughly 80\%, 86\%, and 85\% at a distance of 4\farcs5, and 53\%, 51\%, and 56\% at a distance of 7\farcs5  
for the total, soft, and hard bands, respectively.  
Overall we estimate that 61\%, 65\%, and 65\% of the total, soft and hard band sources have \wise\ counterparts.
The fraction is higher for the hard and soft band catalogs simply because the flux limits will be brighter.  

The distribution of matches for the extended sources looks quite different.  
First, while 12\% of the total band AGN sample had no \wise\ source within
20\arcsec, only 6\% of the cluster candidates had no source within that
distance.  Second, while 2/3 of the AGN candidates had a closest source 
with a red MIR color ($W1-W2 > 0.35$~mag), less than 1/2 of the cluster
candidates do so.  This {\it{does not}} mean that 1/2 of the cluster 
candidates are AGN because of the significant noise in the MIR colors
at fainter magnitudes (see Assef et al.\ 2013).  Third, the match distance distribution of the
sources is broader, as we would expect for extended sources being matched
to a local distribution of galaxies.  Overall, 72\% of the extended sources
are probably associated with a \wise\ source.

The \wise\ color magnitude diagram is a useful diagnostic to separate 
AGN and galaxies.  We focus on the deeper W1 ($3.4\mu$m) and W2 ($4.6\mu$m).
In these Vega magnitudes, stars and early-type galaxies tend to be relatively
blue because their MIR spectral energy distributions (SEDs) are dominated
by the Rayleigh-Jeans tail of the emission by stars.  Quasars have much
flatter SEDs and so appear red (e.g., \citealt{Stern2005}, \citealt{assef13},
\citealt{Chung2014}).  This is true as long as there is significant emission
from the quasar accretion disk (at most redshifts, $z \gs 1$, it is not dust emission),
and red, MIR selected quasars generally prove to be broad line, Type~I
AGN (see Kochanek et al.\ 2012).  X-ray sources extend to bluer MIR 
colors, indicating that the disk emission does not dominate over that of
the host (e.g. \citealt{Gorjian2008}, \citealt{Chung2014}), and these sources will more likely
be Type~II AGN.  
We note that the red and blue MIR colors are opposite to the usual AGN terminology, where
MIR-blue AGN correspond to absorbed AGN and MIR-red AGN correspond to less absorbed AGN.
For the purposes of the paper we will call sources with \wise\ color, $W1-W2 > 0.35$~mag, MIR-red AGN, and 
those with $W1-W2 \le 0.35$~mag MIR-blue AGN, although the boundary is imperfect.  
Moreover, the \wise\ colors are
noisy for the typical source we consider because most of the matches are
to faint \wise\ sources.  This means that the color is only an indicator
of the nature of the source.  The dependencies of AGN properties on the
\wise\ fluxes are discussed extensively in \citet{assef13}.

\begin{figure}
\epsscale{0.5}
\plotone{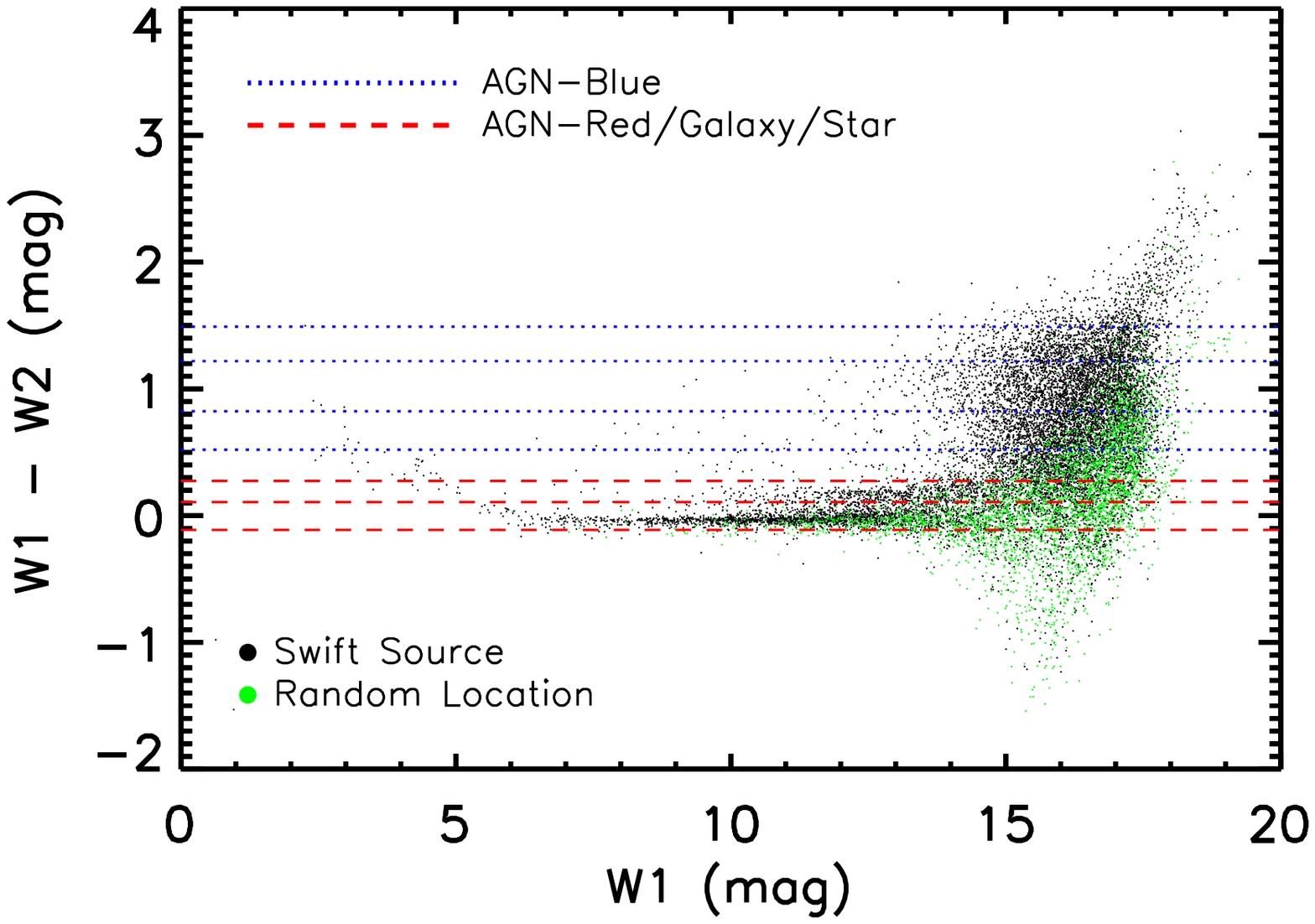}
\plotone{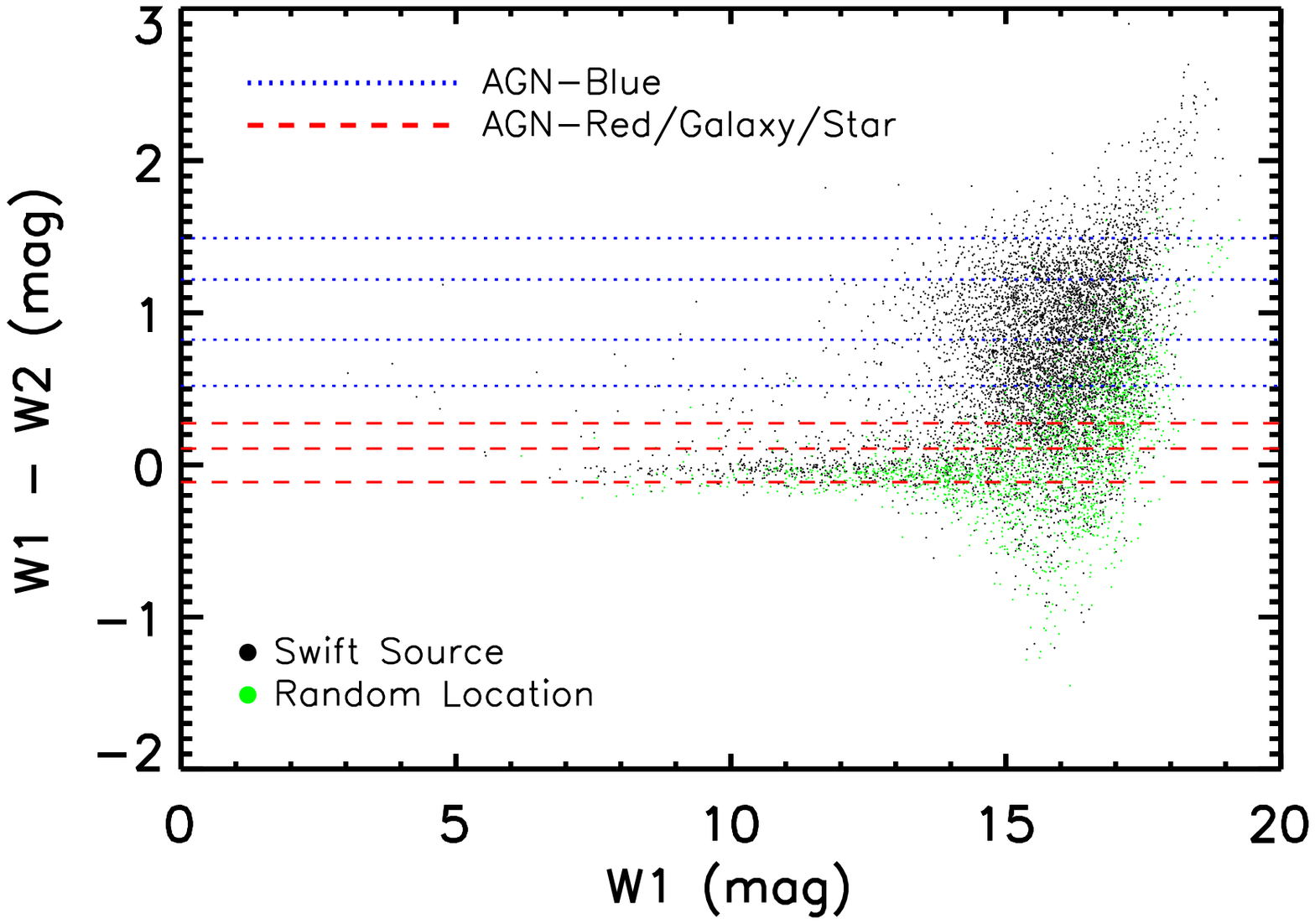}
\plotone{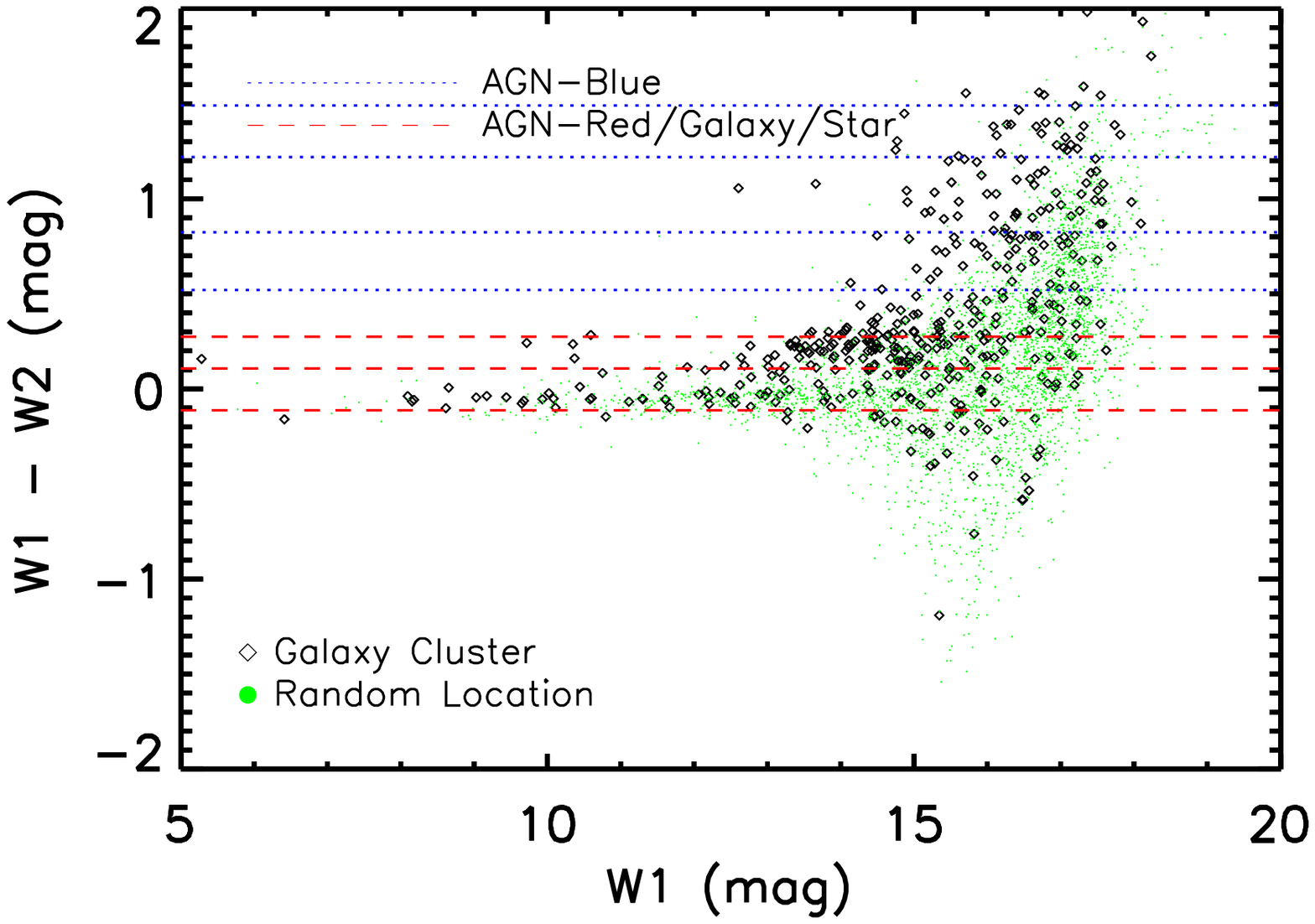}
\caption{\wise\ $W1-W2$ color versus $W1$ magnitude diagrams for \swift\ sources (black dots), random locations (green dots), and Type-I AGN tracks at $z=0.5$, 1.5, 3.5, and 4.5 (blue dotted lines), and early, late, and irregular galaxy tracks (red dashed lines).  
The top and middle panels show the color-magnitude diagrams for point sources detected in the soft and hard X-ray bands, respectively.  The bottom panel shows the color-magnitude diagram for galaxy clusters detected in the soft X-ray band.
\label{fig:cm}}
\end{figure}

Figure~\ref{fig:cm} shows the color magnitude diagrams for the soft and
hard band AGN samples as well as the cluster samples along with the 
typical colors of galaxies and Type~I AGN at various redshifts.  
Here we use only objects with a match distance smaller
than 8\arcsec.  As expected, the AGN samples tend to have relatively
red MIR colors while the sources matched to clusters have the 
MIR colors of intermediate redshift galaxies.  If we use a 
two-dimensional Kolmogorov--Smirnov test to compare the distributions
of AGN and extended sources, there is a null probability of $4\times 10^{-26}$
that the two samples are drawn from the same distribution, further 
showing that the extended X-ray sources are not significantly contaminated
by AGN.

\begin{figure}
\epsscale{1}
\plotone{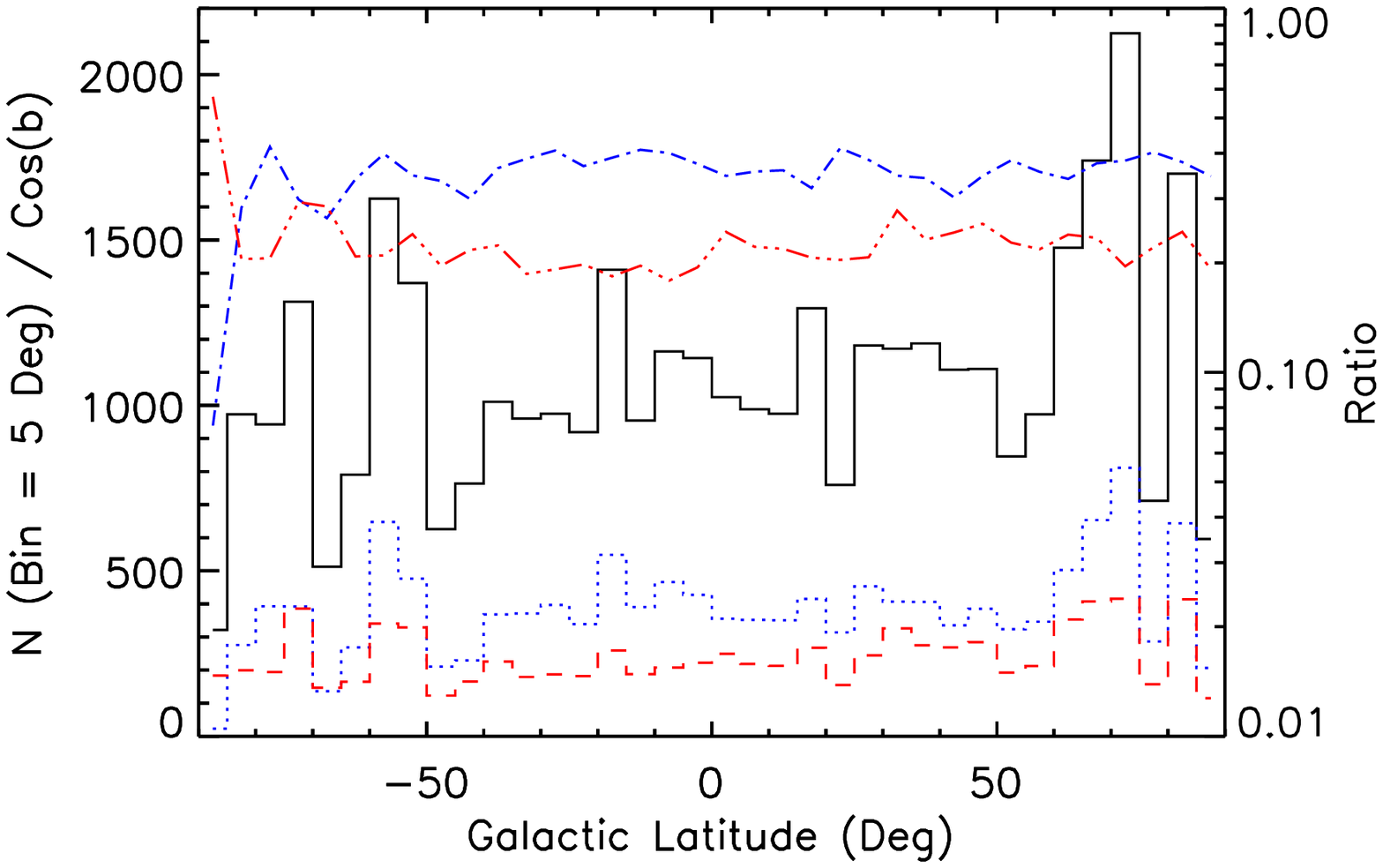}
\caption{ 
   Galactic latitude ($b$) histograms for the total band X-ray sources divided by $\cos{b}$, where flat distributions are expected for sources distributed randomly on the sky.  The black, solid histogram is for all detected sources, the blue, dotted histogram is for the MIR-red AGN, and the red, dashed histogram is for the MIR-blue AGN.  The dotted (blue) and dashed (red) lines show the ratios of MIR-red and MIR-blue AGN to all sources in each bin, respectively.  The number ratios (blue dot-dashed and red dot-dot-dot-dashed lines) do not show a peak near the Galactic plane, indicating that the catalogs contain few stellar sources.  \label{fig:gb}}
\end{figure}

MIR-blue AGN have colors similar to stars as well as galaxies.  However,
Figure~\ref{fig:gb} shows that there is no significant Galactic latitude dependence to either the 
number density of sources or the ratios of the numbers of red and blue MIR sources.  
This strongly suggests that the level of stellar contamination is small.

\section{Comparison with Other \swift\ XRT Catalogs\label{sec:com}}
There are several other \swift\ XRT source catalogs in existence.  We can characterize them by the type of sources (point/extended), 
years of data used (5--7), fields analyzed (GRB or all), and detection method.
They are Puccetti et al.\ (2011, point, 5 year, GRB, \verb+XIMAGE+), D'Elia et al.\ (2013, point, 7 year, all, \verb+detect+), Evans et al.\ (2014, point, 7 year, all, similar to the 2XMM catalog of Watson et al.\ 2009), and Tundo et al.\ (2012, extended, 6 year, GRB, \verb+wavdetect+ plus extended source analysis).
Here, we compare them with our catalogs.

\begin{figure}
\epsscale{0.8}
\plotone{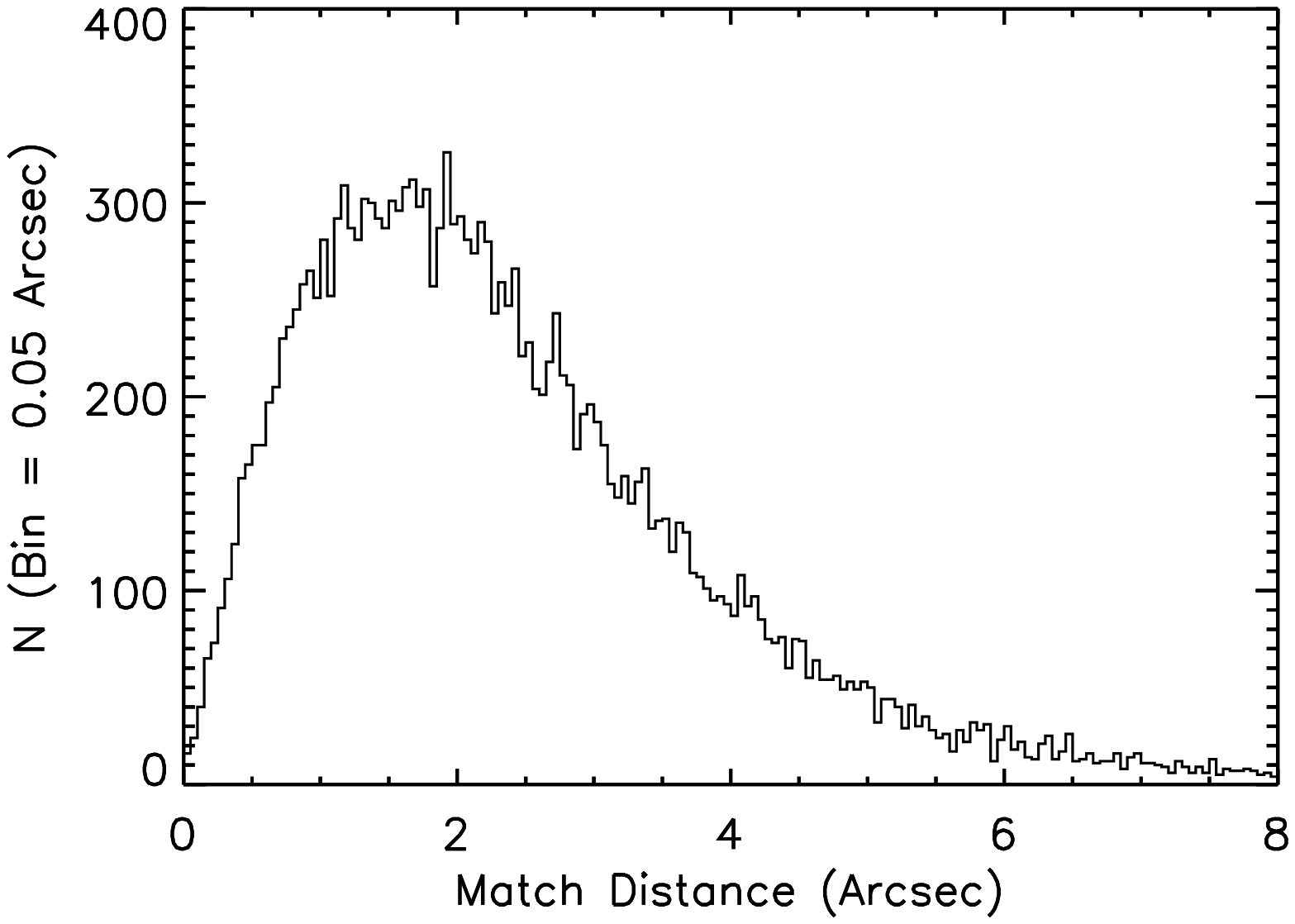}
\plotone{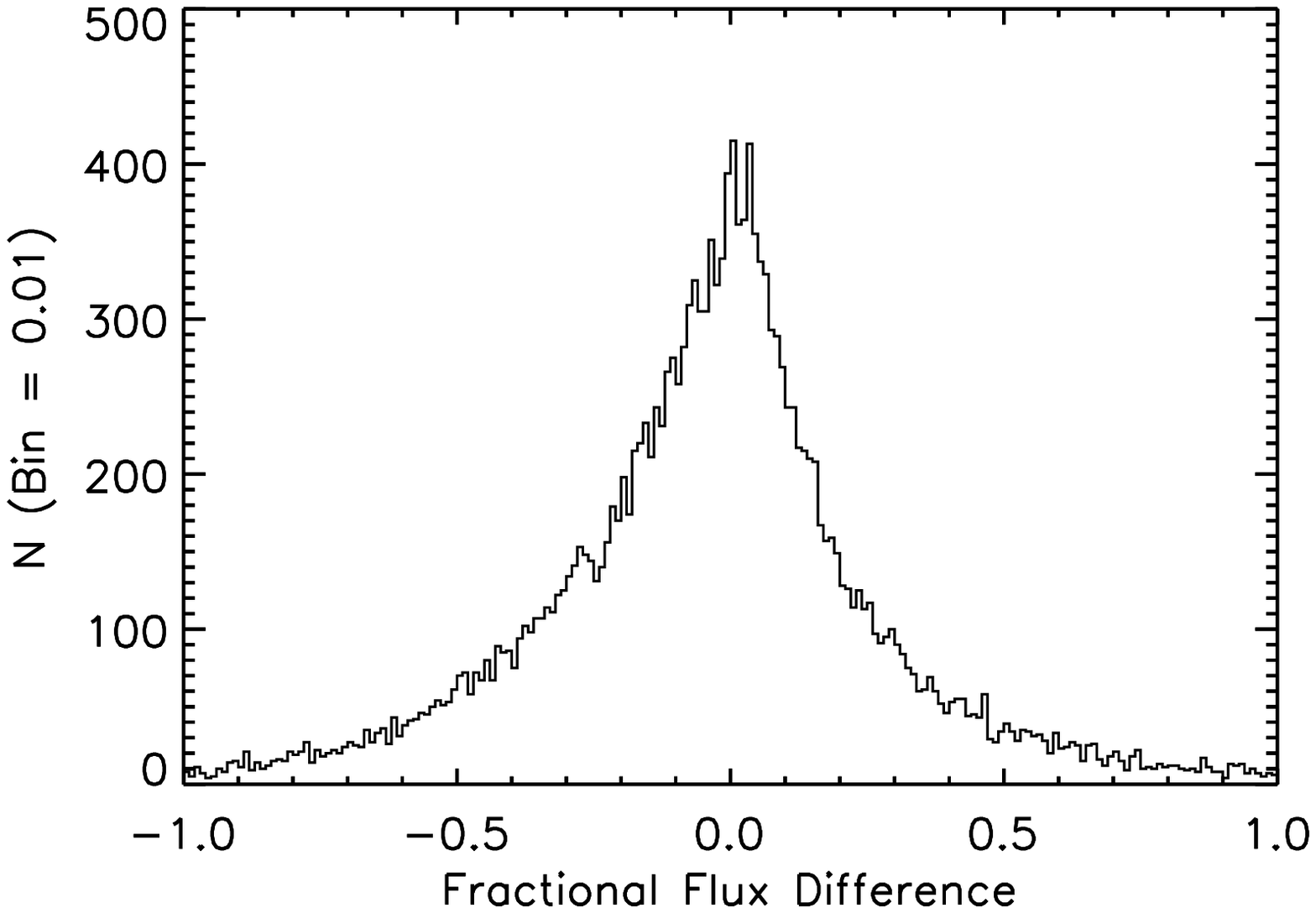}
\caption{Source match distance distribution (top panel) and fractional total flux difference distribution (0.5--8~keV, bottom panel) between the AGN catalog of this paper and \citet{evans14}.  \label{fig:ev}}
\end{figure}

\begin{figure}
\epsscale{0.8}
\plotone{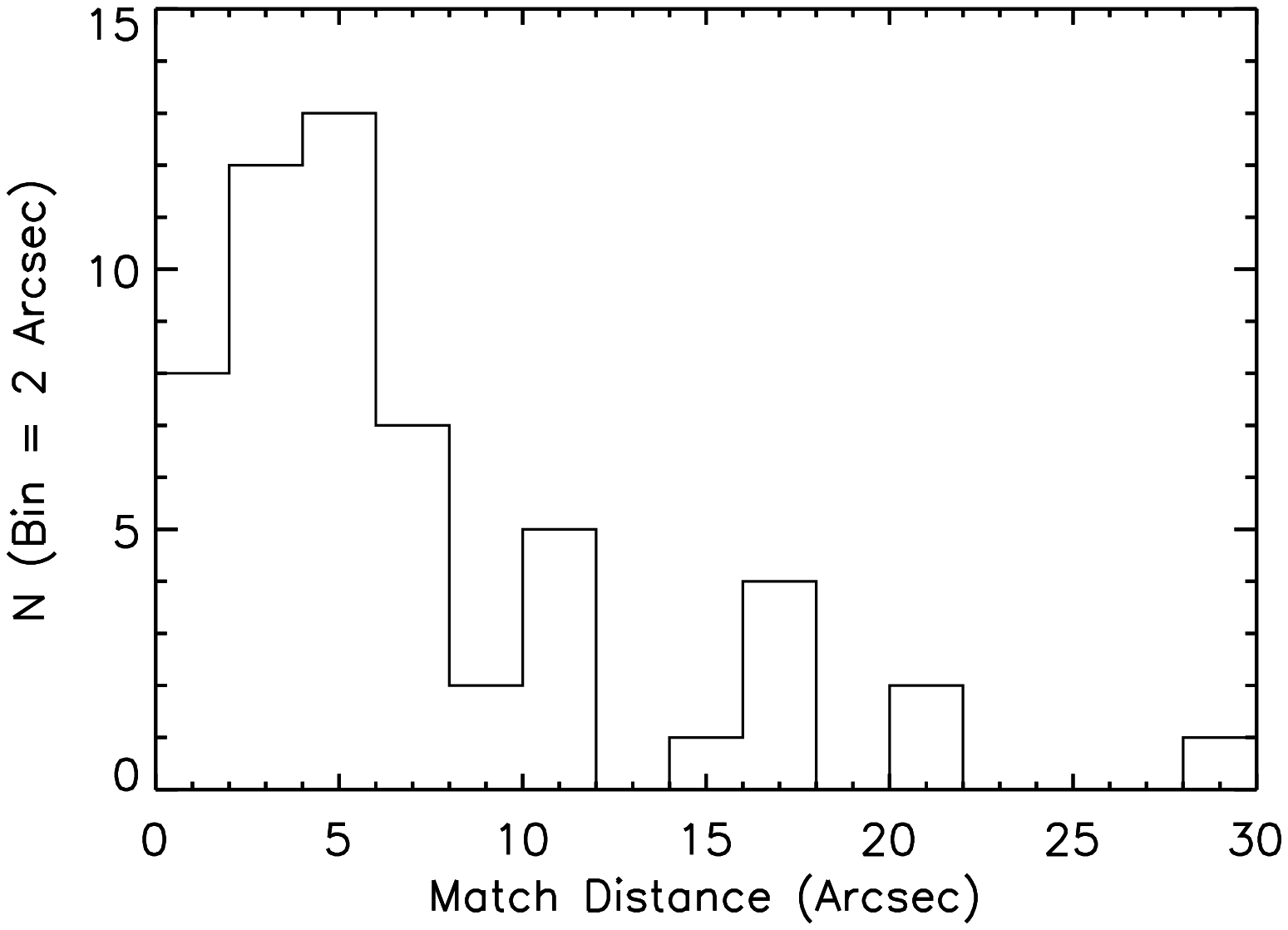}
\plotone{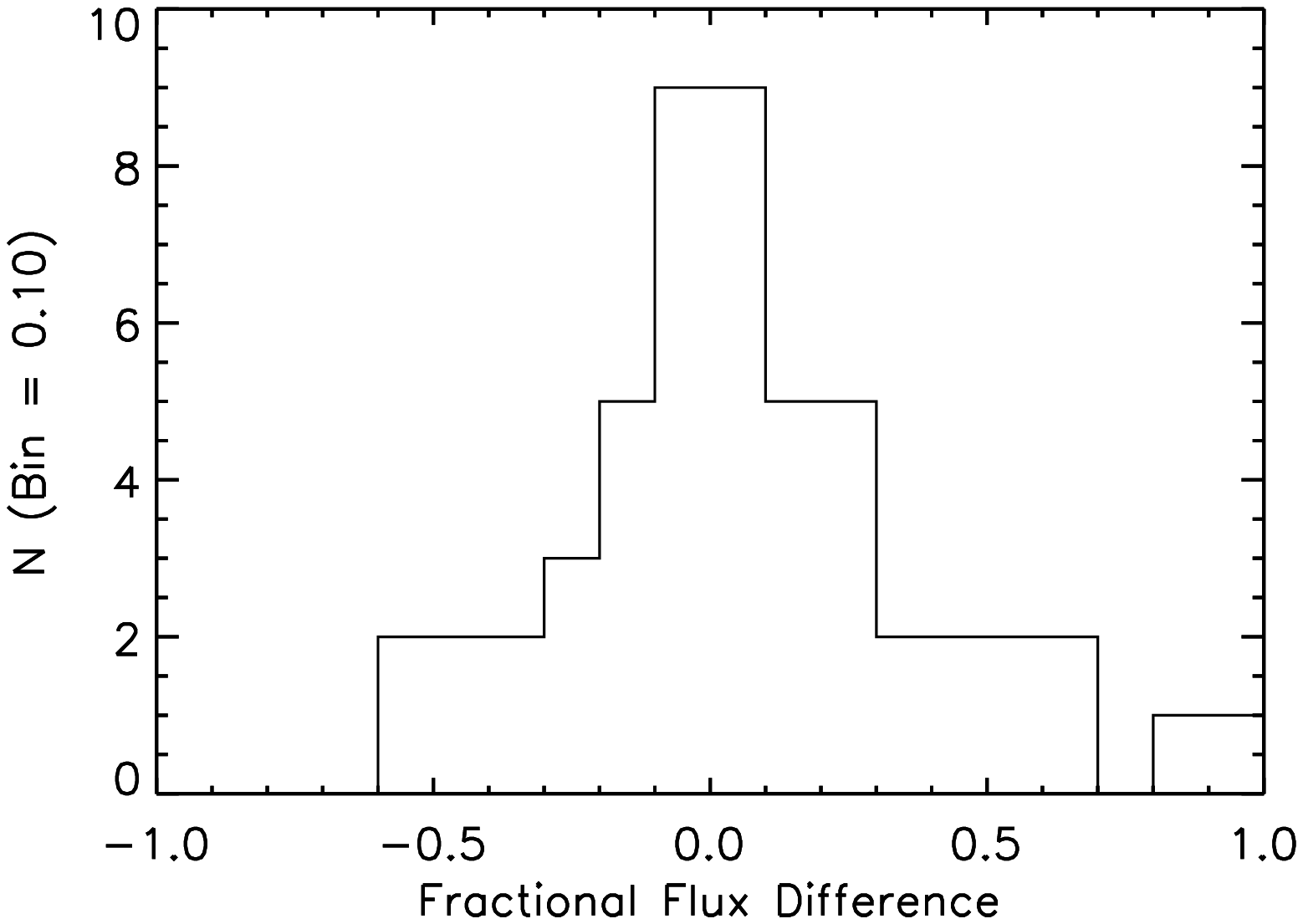}
\caption{Source match distance distribution (top panel) and fractional soft flux difference distribution (0.5--2~keV, bottom panel) between the cluster catalog of this paper and \citet{tundo12}.  \label{fig:tun}}
\end{figure}

\begin{figure}
\epsscale{0.8}
\plotone{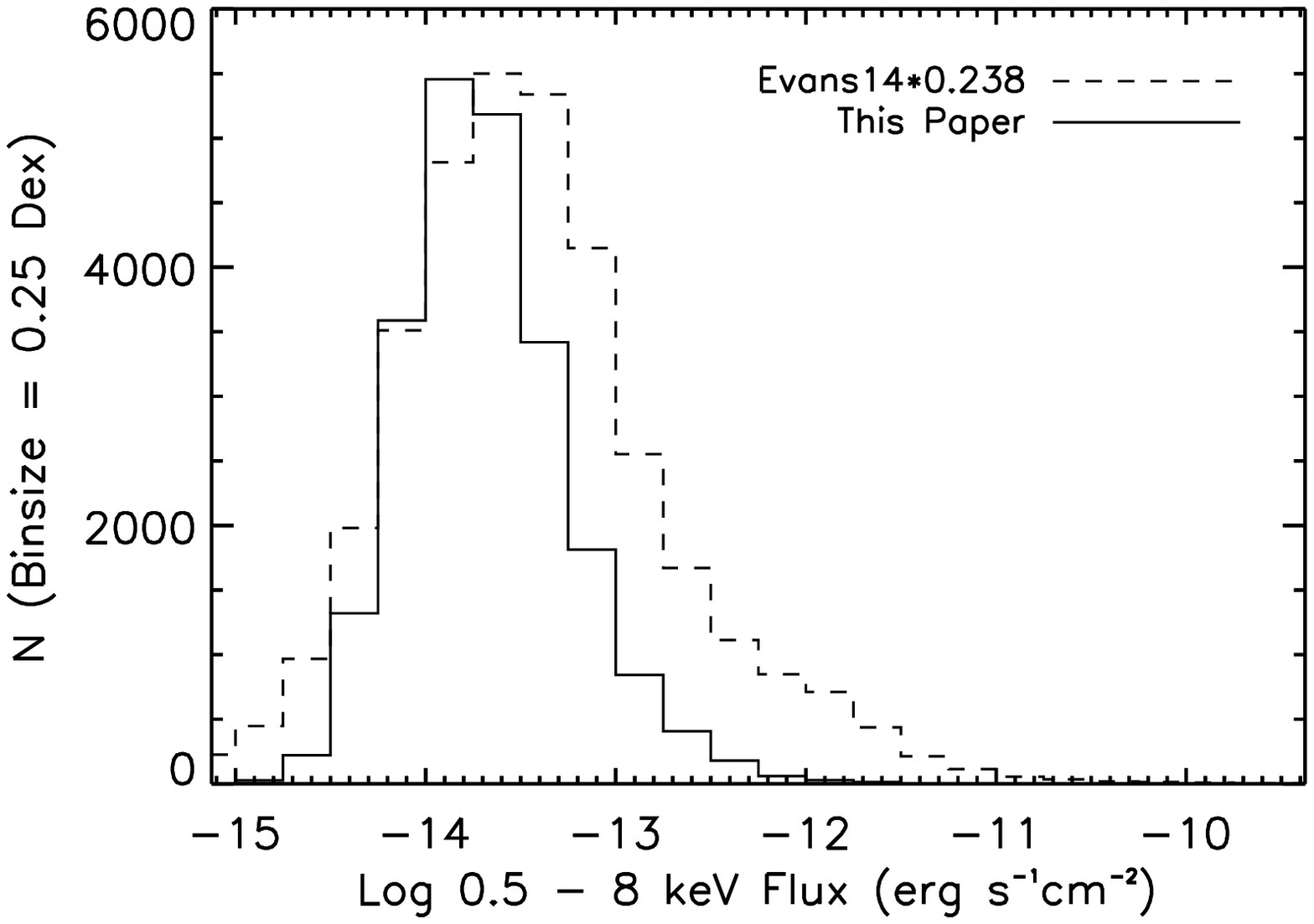}
\plotone{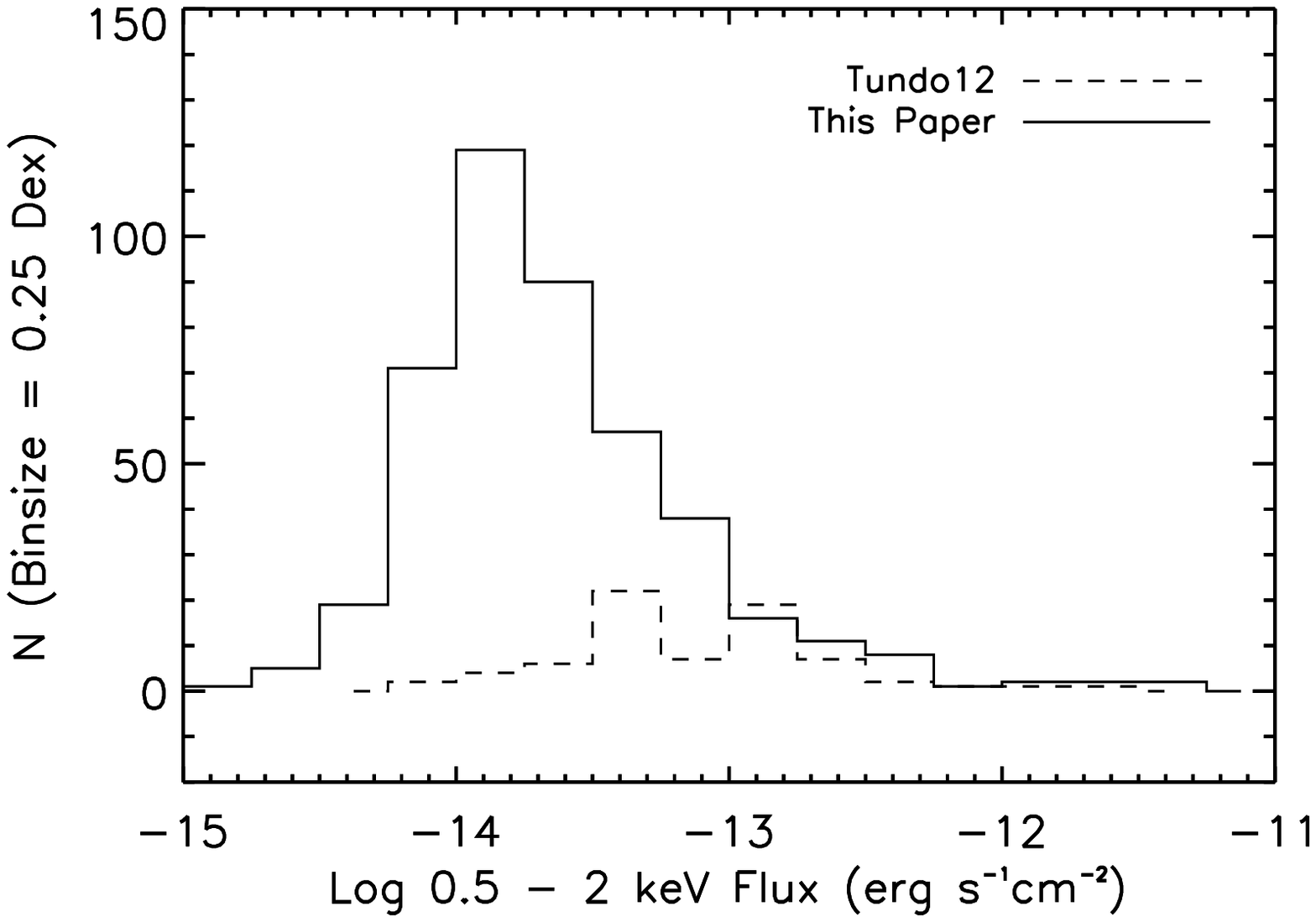}
\caption{Comparisons of the source flux distributions to the AGN catalog of \citet{evans14} normalized by a factor of 0.238 (top panel) and to the cluster catalog of \citet{tundo12} (bottom panel).  \label{fig:fhist}}
\end{figure}

We matched the four point source catalogs to each other using a 8\arcsec\ match distance and the two extended catalogs using a 30\arcsec\ match distance and calculated the match distance distributions.  We also calculated the fractional flux differences, $(f_{cat2} - f_{cat1})/f_{cat1}$, of the matching sources in the 0.5--8~keV band for point sources and the 0.5--2~keV band for extended sources.  
We list the number of matches, the median match distances, the median and standard deviation of the fractional flux differences in Table~\ref{tab:com}, and show the histograms of the matching distances and fractional flux differences between this paper and \citet{evans14} and between this paper and \citet{tundo12} in Figures~\ref{fig:ev} and \ref{fig:tun}, respectively.
In general the source positions are quite consistent between the different catalogs, with median position offsets of \ls 2\arcsec\ for point sources and 5\sarc2 between the two extended catalogs. The $\Delta$RA and $\Delta$Dec distributions are all centered on zero.
Since the fractional flux difference distributions have long tails, we calculated the statistical properties of the distributions within the range of $-1$ to 1.
The fluxes measured from the different catalogs have larger variations, with the standard deviations of the fractional flux differences ranging from 0.15 to 0.39, where the smallest is between the cluster catalog from this paper and \citet{tundo12} and the largest is between the point source catalogs of \citet{delia13} and \citet{evans14}.  The \citet{delia13} catalog also has the largest median offset in the fractional flux difference when matching with other catalogs.
Broadly speaking, our point and cluster catalogs have position and flux measurements consistent with these other analyses.
We compare the flux distributions between our sources and those in \citet{evans14} and \citet{tundo12} for point and extended sources, respectively, in Figure~\ref{fig:fhist}.  For point sources, \citet{evans14} analyzed a much larger amount of \swift\ data and detected a total number of sources 6.7 times our total number.  However, our peak flux distribution is fainter than the sample of \citet{evans14} by 0.5~dex, because GRB fields are in general the deeper fields in the \swift-XRT data set.
For extended sources, at the bright flux end, $>10^{-13}~\flux$, our sample and that of \citet{tundo12} have consistent flux distributions both in the shape and normalization; \citet{tundo12} only reported a tail of extended sources with fluxes below $10^{-13}~\flux$.

\section{Source Number Counts\label{sec:logns}}

We calculate the cumulative number counts ($\log{(N>S)}$--$\log{S}$) of galaxy clusters and AGN detected 
in the survey as
\begin{equation}
N(>S) = \sum_{S_i>S} \frac{1}{\Omega_i C_i},
\end{equation}
where $S_i$ is the total flux of the source, $\Omega_i$ is the total solid angle of the survey reaching flux
$S_i$ and $C_i$ is the incompleteness factor for each source.  The survey area as a function of flux is
calculated from the exposure maps (dashed and dotted lines in Figure~\ref{fig:areaint}), where the areas are normalized to 
an effective \nh\ of $5\times10^{20}\cmsq$.  We model the completeness factor for
point sources using Poisson statistics, and we estimate the completeness factor for extended sources
based on our simulations (Section~\ref{sec:ext}, e.g., Figure~\ref{fig:sig}).  

\begin{figure}
\epsscale{1}
\plotone{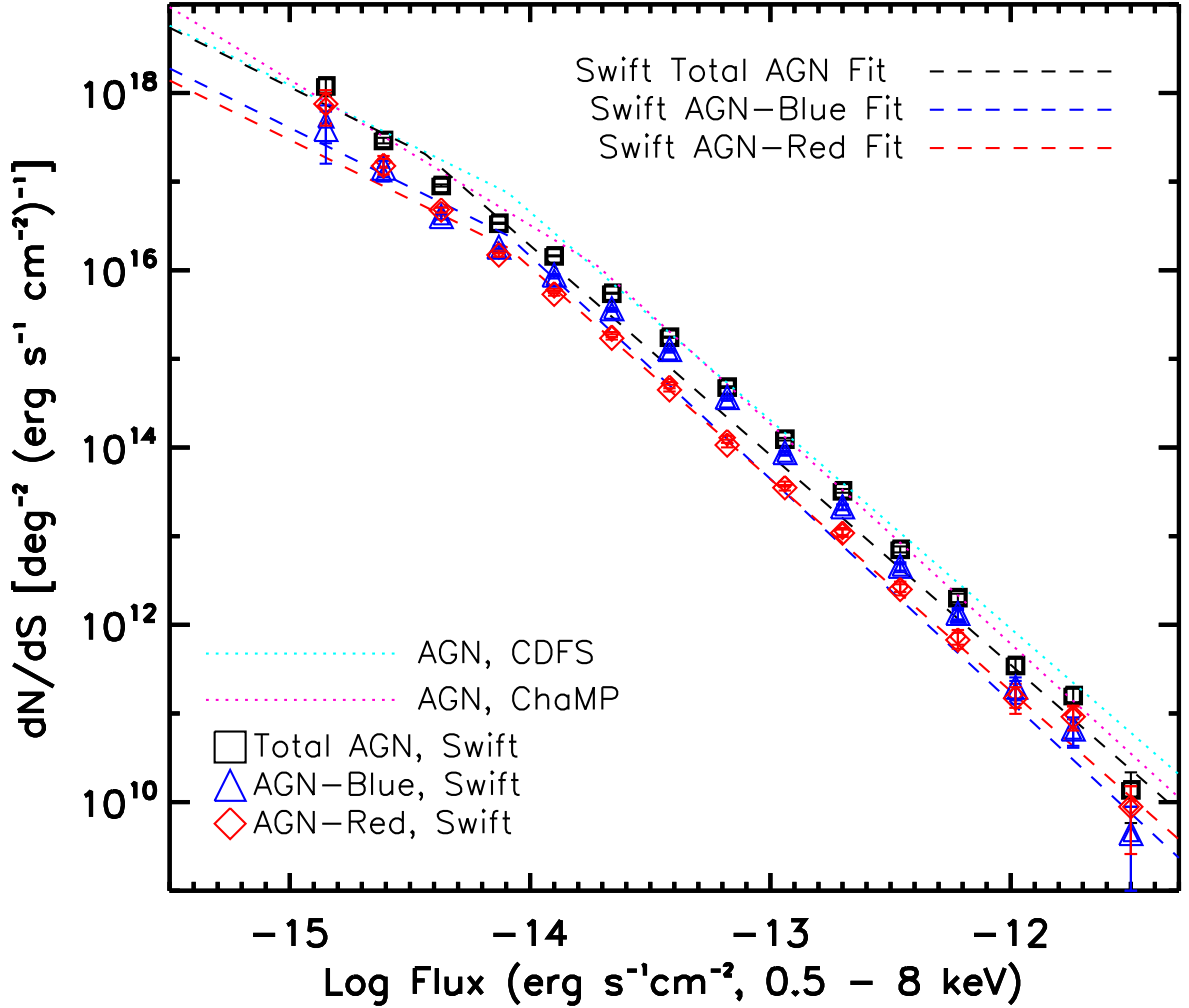}
\vspace{1 cm}
\caption{Differential number counts for the total band ($0.5$--$8$~keV) \swift\ sources for all 
  AGN (black squares), MIR-red AGN (blue triangles, meaning Type~I), and MIR-blue AGN 
  (red diamonds, meaning Type~II).  The larger and smaller symbols 
  (for triangles and diamonds only) are the 5\arcsec\ and 8\arcsec\ \wise\ match results, respectively, where we have corrected the incomplete
\wise\ detection fractions.
  The dotted cyan and magenta lines are the best-fit number counts from the CDF-S and ChaMP.
  The black, blue, and red dashed lines show our fits to the \swift\ number counts for the total, 
  MIR-red, and MIR-blue AGN, respectively.  The MIR-blue AGN represent 50\% of the total band point source detections. \label{fig:agntotal}}
\end{figure}

\begin{figure}
\epsscale{1}
\plotone{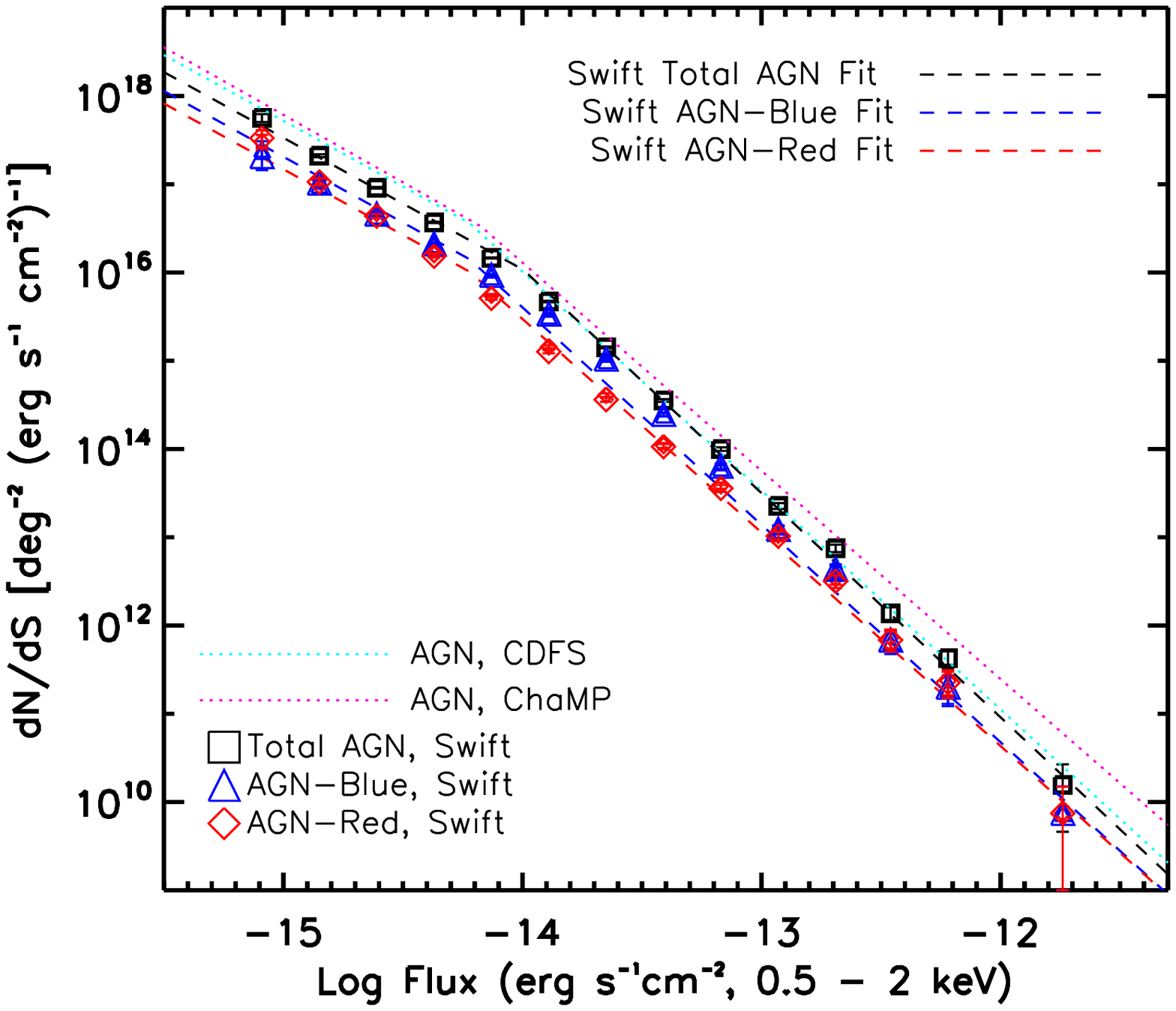}
\vspace{1 cm}
\caption{Differential number counts for the soft band ($0.5$--$2$~keV) \swift\ sources for all 
  AGN (black squares), MIR-red AGN (blue triangles, meaning Type~I), and MIR-blue AGN 
  (red diamonds, meaning Type~II).  The larger and smaller symbols 
  (for triangles and diamonds only) are the 5\arcsec\ and 8\arcsec\ \wise\ match results, respectively, where we have corrected the incomplete
\wise\ detection fractions.
  The dotted cyan and magenta lines are the best-fit number counts from the CDF-S and ChaMP.
  The black, blue, and red dashed lines show our fits to the \swift\ number counts for the total, 
  MIR-red, and MIR-blue AGN, respectively.  The MIR-blue AGN represent 50\% of the soft band point source detections. \label{fig:agnsoft}}
\end{figure}

\begin{figure}
\plotone{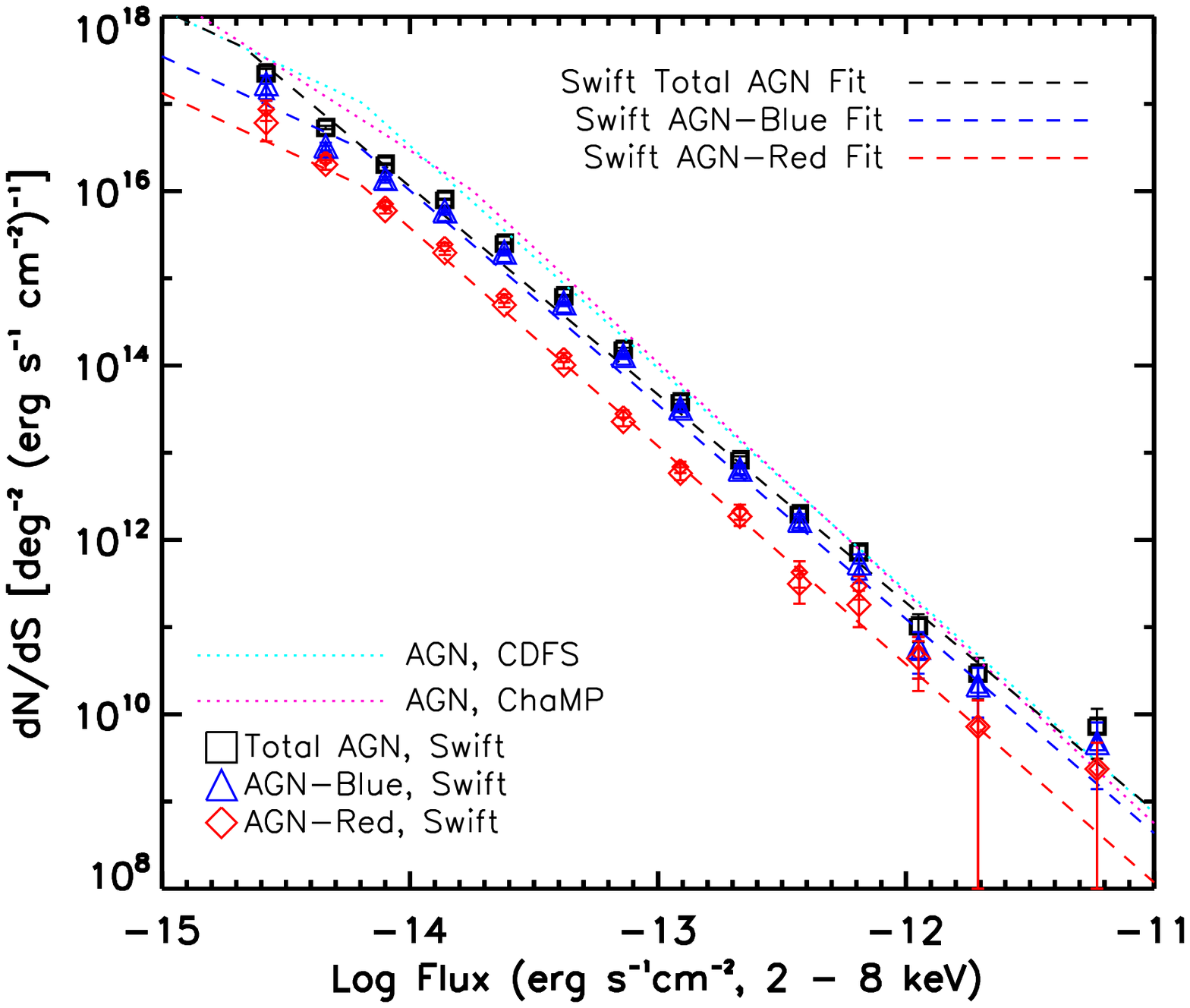}
\vspace{1 cm}
\caption{Differential number counts for the hard band ($2$--$8$~keV) \swift\ survey for all AGN (black squares), 
  MIR-red AGN (blue triangles, meaning Type~I), and MIR-blue AGN (red diamonds, meaning Type~II).
  The larger and smaller symbols (for triangles and diamonds only) are the 5\arcsec\ and 8\arcsec\ \wise\ match 
  results, respectively, where we have corrected the incomplete
\wise\ detection fractions.  The dotted cyan and magenta lines are the best-fit number counts from the CDF-S and ChaMP.
  The black, blue, and red dashed lines are our fits to the \swift\ number counts for the total, MIR-red, and 
   MIR-blue AGN, respectively. The MIR-blue AGN represent 25\% of the hard band point source detections. \label{fig:agnhard}}
\end{figure}

\begin{figure}
\plotone{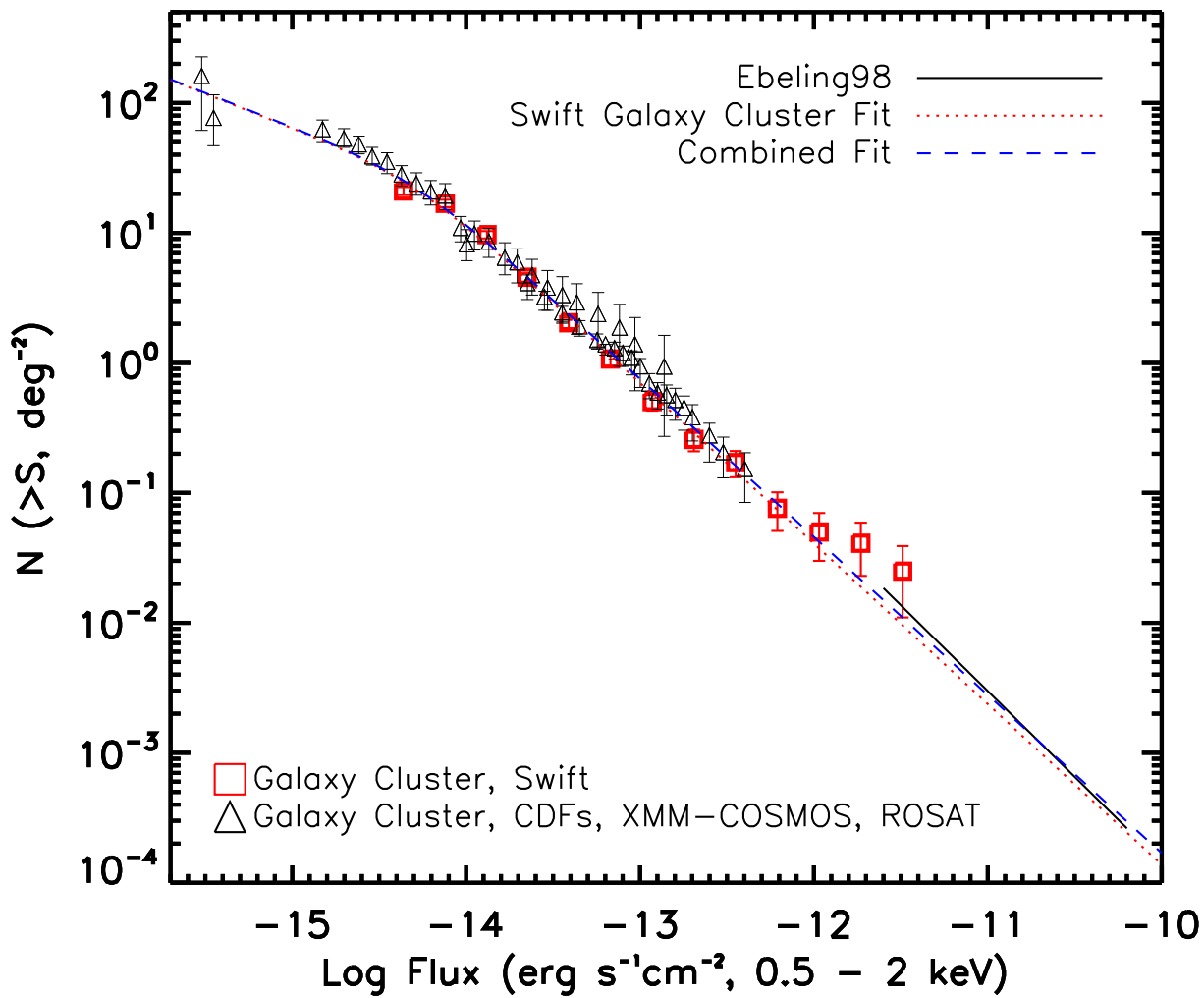}
\vspace{1 cm}
\caption{Cumulative number counts of galaxy clusters in the $0.5$--$2$~keV band from the 
  \swift\ survey (red squares) as compared to results from the XMM-COSMOS, ROSAT, and CDF-S 
  surveys (black triangles and the black solid line).  The red dotted line shows the broken power-law fit to the 
  \swift\ data only, and the blue dashed line shows the fit to the combined data with fluxes lower than $4\times10^{-12}~\flux$. \label{fig:clsoft}}
\end{figure}

Figures~\ref{fig:agntotal}, \ref{fig:agnsoft} and \ref{fig:agnhard} show the differential number counts for AGN in the 
total, soft and hard X-ray bands, respectively, and Figure~\ref{fig:clsoft} shows the cumulative cluster 
number counts in the soft X-ray band.  We show differential number counts for AGN because
we have many sources and this makes the uncertainties in the flux bins independent.  For the 
clusters we show the integral number counts because the differential number counts are
noisy and most previous cluster surveys only provide cumulative number counts.
The \swift\ differential AGN number counts and cumulative cluster number counts are 
provided in Tables~\ref{tab:agndiffsoft}, \ref{tab:agndiffhard}, and \ref{tab:clcumul}.

AGN number counts are generally modeled as a broken power-law with a sharp break, and
Figures~\ref{fig:agntotal}, \ref{fig:agnsoft} and \ref{fig:agnhard} show that our AGN number counts are 
consistent with previous measurements in both the total, soft, and hard X-ray bands. Here
we use the \chandra\ Deep Field South (CDF-S, Lehmer et al.\ 2012) and the 
ChaMP survey (Kim et al.\ 2007) as the comparison samples.  With the large number of sources 
detected in the \swift\ fields, we can place tighter constraints on the slope of the bright 
end of number counts than previous studies.  We fit the \swift\ differential number counts 
($dN/dS$) of AGN using the standard broken power law model, 
\be
dN/dS = \left\{
    \begin{array}{ll}
        K (S/S_{ref})^{-a}                          & (S \le f_b) \\
        K (f_b/S_{ref})^{b-a}(S/S_{ref})^{-b}       & (S > f_b)
    \end{array} \right.
    \label{eq:agn}
\ee
where $S_{ref}$ is a reference flux set to $10^{-14}~\flux$, $f_b$ is the break flux, and 
$K$, $a$ and $b$ are the normalization, and power-law indices at the low and high flux ends, 
respectively.  The parameters of the models are provided in Table~\ref{tab:agnpara}.
We fixed the power-law index for faint sources, $a$, to that measured for the CDF-S 
since the \swift\ number counts do not extend much below the break. 
We measure bright end slopes of $b=2.37\pm0.01$, $2.54\pm0.04$ and $2.39\pm0.14$ 
for the total, soft and hard bands, respectively.  The measurement uncertainties are smaller 
than the corresponding uncertainties from the CDF-S, $b=2.35\pm0.15$, $2.48\pm0.27$ and $2.55\pm0.17$, and 
comparable to the ChaMP measurements of $b=2.48\pm0.05$, $2.36\pm0.05$, and $2.65\pm0.07$.  
Our results agree more with the CDF-S results at the bright fluxes and modestly disagree with the ChaMP results. 

These uncertainties only represent Poisson errors due to the number of sources.  Cosmic variance can be a 
more important source of differences between surveys due to large scale structures.
 Using $10^{-14}~\flux$ in the 0.5--2~keV band as a reference, we estimate that the redshift limit for $L^*_X$ AGN with this flux is $z>4$ because of their rapid positive luminosity evolution \citep{hasinger05}.  
At fixed flux, the survey volume is simply proportional to the survey area, and thus SACS will provide the largest survey volume and the smallest cosmic variance.  Based on the cosmic variance estimator of \citet{ts08}, we estimate that the level of cosmic variance in CDF, ChaMP, and SACS are $\sim$15, 2, and 1\%, respectively.  
Thus, the differences between CDF and SACS can be easily explained by cosmic variance; however, the differences between ChaMP and SACS must originate from other sources of systematic uncertainties, such as the source detection and flux estimation methods.

We can approximate the division into Type I and II AGN using the MIR colors of the
sources with \wise\ matches.  Generally, the objects selected to have red MIR colors
will also show broad line spectra (see Kochanek et al.\ 2012).  To maximize purity, we
keep only sources matched within 5\arcsec\ although we find no significant differences
if we use 8\arcsec.
Here we fix both the faint end slope $a$ and the break flux $f_b$ to the values found
for the CDF-S.  
Not all \swift\ AGN have \wise\ matches, and for simplicity we correct this 
by multiplicatively increasing the number counts by the inverse of the 
matched fractions in each bin.
This is an over-simplification given any correlation between the X-ray and MIR fluxes. 
Comparing the number counts between MIR-red and MIR-blue AGN in the 
total and soft band, we find that there are similar numbers of each, while for the hard
band we find that only about 1/3 are MIR-blue.  This result is somewhat unexpected 
because we expected that the spectra of MIR-blue AGN, corresponding to Type~II AGN, would be harder because of 
the higher intrinsic absorption.  
It is possible that many of the harder, Type II X-ray sources remain mid-IR red, suggesting
      that the total absorption column densities are modest.   Optical (visual) and 
      soft X-ray (keV) opacities are similar for typical dust-to-gas ratios \citep[e.g.,][]{dk09, chen13}, so the
      transition from Type I to Type II as an X-ray source will be associated with
      suppression of the optical signatures of an AGN.  
However, mid-IR dust opacities are over an order of magnitude lower, so there is an extended
      range of intermediate column densities ($\nh \sim 10^{23}~\cmsq$), where absorption
      will produce the optical and X-ray signatures associated with Type II AGN 
      while having little effect on the mid-IR colors.
With the large number of sources detected, we can 
measure the intrinsic absorption by stacking the source events based on the flux bins, 
and we plan to explore this further in follow-up studies.

For the cluster number counts, we compare the \swift\ cumulative number counts with the 
composite cumulative number counts from Rosati et al.\ (2002) which combines the
\rosat, CDF-S and \xmm-COSMOS field (Finoguenov et al.\ 2007) results for the soft 
X-ray band (0.5--2~keV).  The number counts of our extended source catalog match
these well, strongly indicating that it is dominated by real clusters. 
We fit the cumulative cluster number counts with a smooth broken power-law model,
\be
N(>S) = K \frac{(S/f_b)^{-a}}{(1+(S/f_b)^c)^{(b-a)/c}}, 
\label{eq:cl}
\ee
where $f_b$ is the break flux, $c \equiv 2 $ is a fixed ``smoothness'',  $K$ sets the normalization
and $a$, $b$ are the power-law indices at faint and bright fluxes.  We first fit the \swift\ data 
only and then all available data in Figure~\ref{fig:clsoft}, where the fitting results are listed 
in Table~\ref{tab:clpara}.  Since the \swift\ constraints do not extend below the break of the 
relation, we fixed the break location and the power law index at faint fluxes to be that found
for the global fit using all the data.  The two fits are consistent in normalization and 
the power-law index for bright sources.  We find that the 
power-law index for bright sources is $b=1.25\pm0.04$ and $b=1.24\pm0.01$ using the \swift\ and 
the combined data, respectively. This is significantly different from Euclidean slope of $1.5$,
indicating that there is strong source evolution with redshift, as expected for a late-forming
population like rich galaxy clusters.  
At brighter fluxes than probed by SACS, $> 4\times10^{-12}~\flux$, the number counts steepen to $b=1.32\pm0.05$ \citep{ebeling98}, and the slope reaches the Euclidean value at fluxes above $> 10^{-11}~\flux$ \citep{ebeling98, degrandi99}.
We also estimate the effects of cosmic variance at a reference flux of $10^{-14}~\flux$, and set a redshift limit of $z<1.5$ in these estimates, since clusters are lower redshift objects.
We find that cosmic variance can contribute $\sim$15 and 5\% uncertainties for \xmm-COSMOS and SACS surveys, respectively, which are consistent with the differences between the number count measurements of the two surveys.

\section{Discussion}

\swift\ observations of GRB fields form an excellent serendipitous soft X-ray survey 
with a medium depth and wide sky coverage.  For example, the total number of sources detected in
our survey ($\sim 22,000$) is larger than the RASS sample ($\sim 18,000$, Voges et al.\ 1999). 
The large number of sources enables us to more accurately measure source number counts, 
especially for bright sources and with few systematics due to cosmic variance.
Since GRBs are not correlated with other sources, there is little selection bias for 
extragalactic objects.  This is very different from serendipitous surveys based
on \chandra\ and \xmm\ where many of the fields are selected to study either clusters
or AGN.  In addition, the relative uniformity of the \swift-XRT PSF across its field of view
allows for much more uniform source detection and characterization than those for \rosat, \chandra, or \xmm,
where the PSF strongly or moderately depends on the off-axis angle.

The next step is to study these sources in detail using multi-wavelength data.  In 
this paper, we matched the sources to the \wise\ all-sky catalog and found 15,100 matches 
of the X-ray sources.  We can use the \wise\ $W1-W2$ color as a proxy for separating
the sources into Type I (MIR-red sources) and Type II (MIR-blue sources) since
red MIR sources are typically found to be broad line quasars (see Kochanek et al.\ 2012). 
In the total and soft band about 50\% of the AGN are Type~II, similar to the distributions
seen in \cite{Chung2014}. 
We plan to extend these comparisons using the $\sim 9000$~deg$^2$ coverage
of the SDSS.  However, more complete identifications across the full sky will require
the new generation of optical surveys such as PanSTARRS, DES, and LSST.

The depth and areal coverage of the \swift\ serendipitous survey is well suited for 
galaxy cluster surveys outside the local universe, reaching $z\sim 1$ for massive
clusters. In this paper we have identified 442 cluster candidates as extended X-ray
sources.  Their measured number counts are consistent with previous studies and
they show few matches to red MIR sources, so the contamination by AGN is small.  
Compared to optical cluster surveys where the numbers of candidates can exceed
50,000 \citep{hao10}, the total numbers of X-ray selected clusters are relatively small.
For example, Piffaretti et al.\ (2011) compiled a meta X-ray cluster catalog with 
1,743 clusters.  Our \swift\ cluster sample significantly expands the number of 
X-ray selected clusters, and it is one of the largest uniformly selected cluster samples.
Based on the models of cluster mass functions as a function of redshift (e.g., Tinker et al.\ 2008) 
and the flux limits of our survey, we expect that the peak of our cluster redshift distribution
is at $z\sim 0.5$. In Griffin et al.\ (2015), we use the SDSS survey data
to identify roughly half of clusters in the SDSS regions as $z<0.5$ clusters, 
consistent with the model prediction.  We expect the remainder are higher redshift
clusters, but the SDSS data is too shallow to provide robust photometric redshifts. 
We have started dedicated optical follow-up programs to further measure the photometric 
redshift of the \swift\ clusters, and we expect to significantly increase the number 
of X-ray selected clusters at $z>0.5$.  The sample could be expanded further by
relaxing the 3$\sigma$ selection limit in regions where deep optical imaging data
is available to clearly identify optical over-densities or red galaxy sequences. 

Since the \swift-XRT is still routinely observing GRB afterglows, the sample size
will continue to increase.   However, the current analysis already uses $\sim 9$~yr 
of the data through mid-2013, so the sample size is unlikely to double and the
final available area will likely be 160--200~deg$^2$ with equivalent on-axis
exposure times exceeding 10~ks.  \swift\ has spent a significant amount of observing 
time on sources other than GRBs, and using all these observations could increase
the sample by a factor of 5 at the price of introducing significantly more complex
selection effects.  Future dedicated wide-field X-ray surveys will greatly improve 
the statistics of the X-ray source properties.  For example, \emph{eROSITA} plans 
to perform three surveys (all-sky, medium, and deep) within the next 5--10 years.  
The current \swift\ serendipitous GRB field data has one third the sky coverage of 
the designed \emph{eROSITA} deep fields and will eventually reach one half to 
two-thirds the size.  This makes our \swift\ survey an excellent prototype
for the wide-field X-ray surveys of the future. 

\acknowledgements We thank the anonymous referee for helpful comments.  We acknowledge the financial support from the NASA ADAP program NNX11AD09G and NSF grant AST-1413056.

\begin{deluxetable}{ccccccccccc}
\rotate
\tabletypesize{\scriptsize}
\tablecolumns{11}
\tablewidth{0pt}
\tablecaption{\swift\ Sources Detected in the Total X-ray (0.2--10~keV) Band \label{tab:stotal}}
\tablehead{
\colhead{Name} & \colhead{RA (J2000)} & \colhead{DEC (J2000)} & \colhead{Count Rate (0.2--10~keV)} & \colhead{Flux (0.5--8~keV)} & \colhead{Exposure} & \colhead{W1} & \colhead{W2} & \colhead{W3} & \colhead{W4} & \colhead{Match R} \\
\colhead{} & \colhead{(deg)} & \colhead{(deg)} & \colhead{($10^{-3}$~cnt~s$^{-1}$)} & \colhead{($10^{-14}~\flux$)} & \colhead{(sec)} & \colhead{(mag)} & \colhead{(mag)} & \colhead{(mag)} & \colhead{(mag)} & \colhead{(arcsec)} 
}
\startdata
SACS J000002.1+443056 &     0.0082$\pm$  0.0003 &    44.5156$\pm$ 0.0002 &     0.900$\pm$  0.106 &      4.20 &     100290 &  15.73 &  15.18 &  11.59 &   8.89 &   3.41 \\
 SACS J000003.8+442911 &     0.0149$\pm$  0.0003 &    44.4864$\pm$ 0.0002 &     0.631$\pm$  0.092 &      2.99 &      95614 &  15.53 &  14.61 &  11.82 &   9.27 &   0.73 \\
 SACS J000003.9+444147 &     0.0155$\pm$  0.0004 &    44.6964$\pm$ 0.0002 &     0.454$\pm$  0.080 &      2.16 &      99454 &  99.00 &  99.00 &  99.00 &  99.00 &  99.00 \\
 SACS J000005.2+443534 &     0.0208$\pm$  0.0003 &    44.5927$\pm$ 0.0002 &     0.911$\pm$  0.103 &      4.21 &     108539 &  16.37 &  15.28 &  12.30 &   9.43 &   3.23 \\
 SACS J000009.6+444130 &     0.0390$\pm$  0.0003 &    44.6916$\pm$ 0.0002 &     0.353$\pm$  0.070 &      1.71 &     103680 &  16.30 &  15.88 &  13.06 &   9.35 &   0.72 \\
 SACS J000010.4-344341 &     0.0427$\pm$  0.0007 &   -34.7280$\pm$ 0.0005 &     0.911$\pm$  0.274 &      3.51 &      14190 &  13.81 &  13.23 &   9.81 &   7.78 &   4.64 \\
 SACS J000014.8-344322 &     0.0613$\pm$  0.0007 &   -34.7227$\pm$ 0.0003 &     0.407$\pm$  0.151 &      1.64 &      21921 &  16.08 &  15.45 &  12.87 &   9.18 &   3.74 \\
 SACS J000018.1-344836 &     0.0748$\pm$  0.0004 &   -34.8099$\pm$ 0.0002 &     1.766$\pm$  0.296 &      6.67 &      22194 &  16.29 &  15.81 &  12.44 &   9.03 &   2.02 \\
 SACS J000019.4-345317 &     0.0803$\pm$  0.0006 &   -34.8878$\pm$ 0.0003 &     0.564$\pm$  0.209 &      2.28 &      15237 &  17.50 &  16.60 &  12.34 &   8.72 &   1.87 \\
 SACS J000021.2+444342 &     0.0877$\pm$  0.0003 &    44.7283$\pm$ 0.0002 &     0.797$\pm$  0.104 &      3.72 &      94366 &  12.35 &  12.37 &  12.07 &   8.81 &   3.64 
\enddata
\tablecomments{Table~\ref{tab:stotal} is published in its entirety in the electronic edition of the
\emph{Astrophysical Journal}. A portion is shown here for guidance regarding its form and content.}
\end{deluxetable}

\begin{deluxetable}{ccccccccccc}
\rotate
\tabletypesize{\scriptsize}
\tablecolumns{11}
\tablewidth{0pt}
\tablecaption{\swift\ Sources Detected in the Soft X-Ray (0.5--2~keV) Band \label{tab:ssoft}}
\tablehead{
\colhead{Name} & \colhead{RA (J2000)} & \colhead{DEC (J2000)} & \colhead{Count Rate (0.5--2~keV)} & \colhead{Flux (0.5--2~keV)} & \colhead{Exposure} & \colhead{W1} & \colhead{W2} & \colhead{W3} & \colhead{W4} & \colhead{Match R} \\
\colhead{} & \colhead{(deg)} & \colhead{(deg)} & \colhead{($10^{-3}$~cnt~s$^{-1}$)} & \colhead{($10^{-14}~\flux$)} & \colhead{(sec)} & \colhead{(mag)} & \colhead{(mag)} & \colhead{(mag)} & \colhead{(mag)} & \colhead{(arcsec)} 
}
\startdata
SACS J000002.3+443056 &     0.0088$\pm$  0.0003 &    44.5155$\pm$ 0.0002 &     0.536$\pm$  0.078 &      1.53 &     100372 &  15.73 &  15.18 &  11.59 &   8.89 &   3.22 \\
 SACS J000003.7+442911 &     0.0147$\pm$  0.0004 &    44.4863$\pm$ 0.0002 &     0.374$\pm$  0.068 &      1.09 &      95634 &  15.53 &  14.61 &  11.82 &   9.27 &   1.13 \\
 SACS J000004.1+444147 &     0.0163$\pm$  0.0005 &    44.6965$\pm$ 0.0002 &     0.349$\pm$  0.064 &      1.02 &      99552 &  99.00 &  99.00 &  99.00 &  99.00 &  99.00 \\
 SACS J000005.4+443537 &     0.0216$\pm$  0.0003 &    44.5935$\pm$ 0.0002 &     0.588$\pm$  0.078 &      1.71 &     108672 &  16.37 &  15.28 &  12.30 &   9.43 &   0.99 \\
 SACS J000009.5+444129 &     0.0387$\pm$  0.0005 &    44.6914$\pm$ 0.0003 &     0.258$\pm$  0.056 &      0.74 &     103607 &  16.30 &  15.88 &  13.06 &   9.35 &   0.91 \\
 SACS J000014.2+443105 &     0.0585$\pm$  0.0008 &    44.5181$\pm$ 0.0004 &     0.081$\pm$  0.034 &      0.25 &     102784 &  15.81 &  15.91 &  12.48 &   9.38 &   6.66 \\
 SACS J000018.3-344837 &     0.0756$\pm$  0.0005 &   -34.8100$\pm$ 0.0003 &     0.973$\pm$  0.215 &      2.07 &      22345 &  16.29 &  15.81 &  12.44 &   9.03 &   1.16 \\
 SACS J000021.2+444340 &     0.0874$\pm$  0.0004 &    44.7277$\pm$ 0.0003 &     0.508$\pm$  0.079 &      1.45 &      94517 &  12.35 &  12.37 &  12.07 &   8.81 &   4.31 \\
 SACS J000024.3+444057 &     0.1004$\pm$  0.0004 &    44.6825$\pm$ 0.0004 &     0.063$\pm$  0.029 &      0.21 &     104580 &  16.10 &  17.16 &  13.04 &   9.40 &   4.02 \\
 SACS J000025.0+443636 &     0.1035$\pm$  0.0003 &    44.6099$\pm$ 0.0002 &     0.649$\pm$  0.081 &      1.84 &     113763 &  12.06 &  11.96 &  11.94 &   9.18 &   1.74 
\enddata
\tablecomments{Table~\ref{tab:ssoft} is published in its entirety in the electronic edition of the
\emph{Astrophysical Journal}. A portion is shown here for guidance regarding its form and content.}
\end{deluxetable}

\begin{deluxetable}{ccccccccccc}
\rotate
\tabletypesize{\scriptsize}
\tablecolumns{11}
\tablewidth{0pt}
\tablecaption{\swift\ Sources Detected in the Hard X-ray (2--10~keV) Band \label{tab:shard}}
\tablehead{
\colhead{Name} & \colhead{RA (J2000)} & \colhead{DEC (J2000)} & \colhead{Count Rate (2--10~keV)} & \colhead{Flux (2--8~keV)} & \colhead{Exposure} & \colhead{W1} & \colhead{W2} & \colhead{W3} & \colhead{W4} & \colhead{Match R} \\
\colhead{} & \colhead{(deg)} & \colhead{(deg)} & \colhead{($10^{-3}$~cnt~s$^{-1}$)} & \colhead{($10^{-14}~\flux$)} & \colhead{(sec)} & \colhead{(mag)} & \colhead{(mag)} & \colhead{(mag)} & \colhead{(mag)} & \colhead{(arcsec)} 
}
\startdata
SACS J000002.0+443053 &     0.0075$\pm$  0.0006 &    44.5148$\pm$ 0.0004 &     0.248$\pm$  0.057 &      2.04 &     100065 &  15.73 &  15.18 &  11.59 &   8.89 &   2.56 \\
 SACS J000003.7+442910 &     0.0145$\pm$  0.0003 &    44.4861$\pm$ 0.0003 &     0.219$\pm$  0.052 &      1.89 &      95350 &  15.53 &  14.61 &  11.82 &   9.27 &   2.14 \\
 SACS J000005.0+443532 &     0.0200$\pm$  0.0004 &    44.5922$\pm$ 0.0003 &     0.245$\pm$  0.054 &      2.01 &     108319 &  16.37 &  15.28 &  12.30 &   9.43 &   5.53 \\
 SACS J000010.3-344341 &     0.0424$\pm$  0.0004 &   -34.7279$\pm$ 0.0006 &     0.467$\pm$  0.187 &      4.20 &      14126 &  13.81 &  13.23 &   9.81 &   7.78 &   4.26 \\
 SACS J000014.9-344321 &     0.0616$\pm$  0.0007 &   -34.7224$\pm$ 0.0002 &     0.416$\pm$  0.144 &      3.59 &      21939 &  16.08 &  15.45 &  12.87 &   9.18 &   3.46 \\
 SACS J000018.1-344837 &     0.0748$\pm$  0.0008 &   -34.8101$\pm$ 0.0004 &     0.275$\pm$  0.119 &      2.37 &      22288 &  16.29 &  15.81 &  12.44 &   9.03 &   1.68 \\
 SACS J000021.1+444339 &     0.0870$\pm$  0.0005 &    44.7274$\pm$ 0.0003 &     0.193$\pm$  0.051 &      1.65 &      94927 &  12.35 &  12.37 &  12.07 &   8.81 &   4.88 \\
 SACS J000022.8+443004 &     0.0942$\pm$  0.0007 &    44.5011$\pm$ 0.0004 &     0.107$\pm$  0.038 &      0.92 &     105265 &  16.20 &  15.72 &  12.11 &   9.26 &  10.20 \\
 SACS J000027.0+443114 &     0.1115$\pm$  0.0003 &    44.5205$\pm$ 0.0006 &     0.092$\pm$  0.033 &      0.82 &     111900 &  13.81 &  13.51 &  12.87 &   8.93 &   2.35 \\
 SACS J000038.9+443747 &     0.1614$\pm$  0.0005 &    44.6298$\pm$ 0.0003 &     0.308$\pm$  0.061 &      2.53 &     102187 &  14.69 &  14.78 &  13.00 &   8.99 &   3.65 
\enddata
\tablecomments{Table~\ref{tab:shard} is published in its entirety in the electronic edition of the
\emph{Astrophysical Journal}. A portion is shown here for guidance regarding its form and content.}
\end{deluxetable}

\begin{deluxetable}{ccccccccccc}
\rotate
\tabletypesize{\scriptsize}
\tablecolumns{11}
\tablewidth{0pt}
\tablecaption{\swift\ Extended Source Candidates Detected in the Soft X-ray Band (0.5--2~keV) \label{tab:scl}}
\tablehead{
\colhead{Name} & \colhead{RA} & \colhead{DEC} & \colhead{Core} & \colhead{Extent} & \colhead{Count Rate} & \colhead{Flux\tablenotemark{a}} & \colhead{Bkg\tablenotemark{b}} & \colhead{Off-Axis} & \colhead{Exposure} & \colhead{GRB Field} \\
\colhead{} & \colhead{(J2000)} & \colhead{(J2000)} & \colhead{Size} & \colhead{} & \colhead{(0.5--2~keV)} & \colhead{ (0.5--2~keV)} & \colhead{Rate} & \colhead{Angle} & \colhead{} & \colhead{} \\
\colhead{} & \colhead{(deg)} & \colhead{(deg)} & \colhead{(arcsec)} & \colhead{(arcsec)} & \colhead{($10^{-3}$~cnt~s$^{-1}$)} & \colhead{} & \colhead{} & \colhead{(arcmin)} & \colhead{(sec)} & \colhead{} 
}
\startdata
SWCL J000131.7+444414 &     0.3813 &    44.7373 &  15.4 &  50.7 &    1.068$\pm$ 0.159 &   2.67 &   1.97 & 11.3 &      42004 &   grb101225a \\
 SWCL J000251.5-525825 &     0.7139 &   -52.9734 &  13.1 &  22.4 &    0.211$\pm$ 0.026 &   0.41 &   2.32 &  6.6 &     304365 &    grb070110 \\
 SWCL J000314.3-525514 &     0.8088 &   -52.9203 &  32.5 &  83.7 &    2.420$\pm$ 0.088 &   4.69 &   2.22 &  5.6 &     314164 &    grb070110 \\
 SWCL J000323.8-525355 &     0.8483 &   -52.8985 &  45.4 & 104.9 &    3.868$\pm$ 0.113 &   7.49 &   2.23 &  6.2 &     302754 &    grb070110 \\
 SWCL J000344.2-530152 &     0.9332 &   -53.0310 &  41.1 &  62.5 &    1.034$\pm$ 0.058 &   2.00 &   2.23 &  2.5 &     310255 &    grb070110 \\
 SWCL J000755.7-295503 &     1.9822 &   -29.9172 &  13.9 &  46.0 &    0.426$\pm$ 0.087 &   0.81 &   1.57 &  8.6 &      56789 &    grb070611 \\
 SWCL J001004.8+475139 &     2.5196 &    47.8615 &  12.5 &  31.8 &    0.408$\pm$ 0.059 &   1.05 &   1.88 &  8.1 &     115977 &   grb100802a \\
 SWCL J001011.0+475353 &     2.5455 &    47.8986 &  18.6 &  36.5 &    0.467$\pm$ 0.066 &   1.21 &   1.84 & 10.5 &     105736 &   grb100802a \\
 SWCL J001100.4+474827 &     2.7515 &    47.8079 &  11.9 &  31.8 &    0.320$\pm$ 0.064 &   0.83 &   1.82 & 12.1 &      78166 &   grb100802a \\
 SWCL J001338.0-282923 &     3.4084 &   -28.4895 &  18.5 &  48.3 &    0.935$\pm$ 0.196 &   1.81 &   2.19 & 14.9 &      24303 &   grb070721a
\enddata
\tablenotetext{a}{The flux has the unit of $10^{-14}~\flux$.}
\tablenotetext{b}{The background count rate has the unit of $\rm{10^{-4}~cnt~s^{-1}~arcmin^{-2}}$.}
\tablecomments{Table~\ref{tab:scl} is published in its entirety in the electronic edition of the
\emph{Astrophysical Journal}. A portion is shown here for guidance regarding its form and content.}
\end{deluxetable}

\begin{deluxetable}{cccccc}
\tabletypesize{\scriptsize}
\tablecolumns{6}
\tablewidth{0pt}
\tablecaption{Cluster Detection Probability \label{tab:clprob}}
\tablehead{
\colhead{Net } & \colhead{Background} & \multicolumn{4}{c}{Redshift} \\
\colhead{Photons} & \colhead{Level\tablenotemark{a}} & \colhead{$z=0.2$} & \colhead{$z=0.6$} & \colhead{$z=1.0$} & \colhead{$z=1.4$} 
}
\startdata
    20 & 0.0025 & 0.8054 & 0.4725 & 0.3638 & 0.3351 \\ 
    20 & 0.0041 & 0.7885 & 0.4543 & 0.3467 & 0.3178 \\ 
    20 & 0.0068 & 0.7638 & 0.4277 & 0.3270 & 0.3000 \\ 
    20 & 0.0113 & 0.7215 & 0.3923 & 0.2971 & 0.2711 \\ 
    20 & 0.0187 & 0.6572 & 0.3379 & 0.2554 & 0.2343 \\ 
    20 & 0.0308 & 0.5476 & 0.2636 & 0.1971 & 0.1809 \\ 
    20 & 0.0509 & 0.4232 & 0.1923 & 0.1412 & 0.1285 \\ 
    20 & 0.0842 & 0.2818 & 0.1225 & 0.0897 & 0.0833 \\ 
    20 & 0.1392 & 0.1777 & 0.0781 & 0.0599 & 0.0529 \\ 
    20 & 0.2300 & 0.1009 & 0.0460 & 0.0355 & 0.0333 \\ 
\enddata
\tablenotetext{a}{The background level is in units of cnts~pixel$^{-1}$.}
\tablecomments{We assumed $R_c = 0.1$~Mpc and $\beta = 0.6$ for the simulation, corresponding to a total mass of $\sim 2\times10^{14}~\msun$ and close to the mass limit of our survey.  Table~\ref{tab:clprob} is published in its entirety in the electronic edition of the
\emph{Astrophysical Journal}. A portion is shown here for guidance regarding its form and content.}
\end{deluxetable}

\begin{deluxetable}{ccccccc}
\tabletypesize{\scriptsize}
\tablecolumns{6}
\tablewidth{0pt}
\tablecaption{Comparison Between Indepedent \swift\ XRT Catalogs \label{tab:com}}
\tablehead{
\colhead{Catalog} & \colhead{Catalog} & \colhead{Catalog} & \colhead{Number of} & \colhead{Match}         & \colhead{Median Fractional}      & \colhead{Standard}  \\
\colhead{Type}    & \colhead{One}     & \colhead{Two}     & \colhead{Matches}   & \colhead{Dis. (arcsec)} & \colhead{Flux Difference\tablenotemark{a}} & \colhead{Deviation\tablenotemark{b}} 
}
\startdata
AGN     & This Paper & Puccette11 & 8300    & 2.17 & 1.02  &   0.17 \\
AGN     & This Paper & E'Dlia13   & 8988    & 2.00 & 0.53  &   0.21 \\
AGN     & This Paper & Evans14    & 18861   & 2.08 & 0.96  &   0.17 \\
AGN     & Puccette11 & E'Dlia13   & 5182    & 1.54 & 0.55  &   0.21 \\
AGN     & Puccette11 & Evans14    & 9091    & 1.32 & 1.02  &   0.20 \\
AGN     & E'Dlia13   & Evans14    & 79418   & 2.28 & 1.27  &   0.39 \\
Cluster & This Paper & Tundo12    & 55      & 5.20 & 0.97  &   0.15 
\enddata
\tablenotetext{a}{The fractional flux difference is defined as ${\rm (Flux_{cat2} - Flux_{cat1}) / Flux_{cat1}}$.}
\tablenotetext{b}{The standard deviation of the fractional flux difference.}
\tablecomments{The fluxes are calibrated to 0.5--8~keV for AGN and 0.5--2~keV for clusters.}
\end{deluxetable}

\begin{deluxetable}{cccccccc}
\tabletypesize{\scriptsize}
\tablecolumns{8}
\tablewidth{0pt}
\tablecaption{\swift\ Differential AGN Number Counts in the Total X-ray Band (0.5--8~keV) \label{tab:agndifftotal}}
\tablehead{
\colhead{$\log{S}$} & \multicolumn{6}{c}{dN/dS}\tablenotemark{a} & \colhead{Area} \\
\colhead{($\flux$)} & \colhead{\swift-All} & \colhead{uncertainty} & \colhead{\swift-\wise\ Type I} & \colhead{uncertainty} & \colhead{\swift-\wise\ Type II} & \colhead{uncertainty} & \colhead{(Deg$^2$)}
}
\startdata
 -14.85 & 1.20e+18 & 2.84e+17 & 4.97e+17 & 2.26e+17 & 7.23e+17 & 2.69e+17 & 2.802e-2  \\ 
 -14.61 & 2.91e+17 & 2.11e+16 & 1.23e+17 & 2.24e+16 & 1.69e+17 & 2.56e+16 & 6.818e-1  \\ 
 -14.37 & 9.05e+16 & 2.62e+15 & 4.02e+16 & 2.57e+15 & 4.84e+16 & 2.77e+15 & 7.042e+0  \\ 
 -14.13 & 3.39e+16 & 5.90e+14 & 1.69e+16 & 5.71e+14 & 1.64e+16 & 5.73e+14 & 2.776e+1  \\ 
 -13.90 & 1.45e+16 & 1.98e+14 & 8.27e+15 & 1.97e+14 & 6.01e+15 & 1.70e+14 & 5.742e+1  \\ 
 -13.66 & 5.50e+15 & 7.69e+13 & 3.47e+15 & 7.40e+13 & 1.95e+15 & 5.60e+13 & 8.074e+1  \\ 
 -13.42 & 1.75e+15 & 2.97e+13 & 1.20e+15 & 2.83e+13 & 5.29e+14 & 1.90e+13 & 9.756e+1  \\ 
 -13.18 & 4.80e+14 & 1.11e+13 & 3.46e+14 & 1.04e+13 & 1.29e+14 & 6.46e+12 & 1.080e+2  \\ 
 -12.94 & 1.24e+14 & 4.13e+12 & 8.46e+13 & 3.73e+12 & 3.87e+13 & 2.53e+12 & 1.150e+2  \\ 
 -12.70 & 3.24e+13 & 1.57e+12 & 2.09e+13 & 1.39e+12 & 1.13e+13 & 1.02e+12 & 1.191e+2  \\ 
 -12.46 & 7.10e+12 & 5.51e+11 & 4.38e+12 & 4.59e+11 & 2.70e+12 & 3.61e+11 & 1.214e+2  \\ 
 -12.22 & 2.02e+12 & 2.22e+11 & 1.28e+12 & 1.83e+11 & 7.35e+11 & 1.39e+11 & 1.227e+2  \\ 
 -11.98 & 3.47e+11 & 6.95e+10 & 1.80e+11 & 5.21e+10 & 1.65e+11 & 4.99e+10 & 1.239e+2  \\ 
 -11.74 & 1.59e+11 & 3.56e+10 & 6.34e+10 & 2.24e+10 & 9.51e+10 & 2.74e+10 & 1.247e+2  \\ 
 -11.50 & 1.37e+10 & 7.92e+09 & 4.43e+09 & 4.43e+09 & 8.86e+09 & 6.26e+09 & 1.251e+2
\enddata
\tablenotetext{a}{$dN/dS$ has the unit of Deg$^{-2}~(\flux)^{-1}$.}
\end{deluxetable}

\begin{deluxetable}{cccccccc}
\tabletypesize{\scriptsize}
\tablecolumns{8}
\tablewidth{0pt}
\tablecaption{\swift\ Differential AGN Number Counts in the Soft X-ray Band (0.5--2~keV) \label{tab:agndiffsoft}}
\tablehead{
\colhead{$\log{S}$} & \multicolumn{6}{c}{dN/dS}\tablenotemark{a} & \colhead{Area} \\
\colhead{($\flux$)} & \colhead{\swift-All} & \colhead{uncertainty} & \colhead{\swift-\wise\ Type I} & \colhead{uncertainty} & \colhead{\swift-\wise\ Type II} & \colhead{uncertainty} & \colhead{(Deg$^2$)}
}
\startdata
 -15.09 & 5.66e+17 & 5.84e+16 & 2.50e+17 & 5.71e+16 & 3.02e+17 & 5.93e+16 & 5.10e-1 \\  
 -14.85 & 2.10e+17 & 7.71e+15 & 9.84e+16 & 8.20e+15 & 1.13e+17 & 8.65e+15 & 5.79e+0 \\  
 -14.61 & 9.12e+16 & 1.89e+15 & 4.39e+16 & 1.78e+15 & 4.67e+16 & 1.86e+15 & 2.23e+1 \\  
 -14.37 & 3.69e+16 & 5.93e+14 & 2.07e+16 & 5.78e+14 & 1.60e+16 & 5.15e+14 & 4.92e+1 \\  
 -14.13 & 1.46e+16 & 2.27e+14 & 8.99e+15 & 2.15e+14 & 5.54e+15 & 1.69e+14 & 7.26e+1 \\  
 -13.89 & 4.69e+15 & 8.64e+13 & 3.23e+15 & 8.14e+13 & 1.44e+15 & 5.46e+13 & 9.04e+1 \\  
 -13.65 & 1.41e+15 & 3.35e+13 & 1.00e+15 & 3.04e+13 & 4.07e+14 & 1.94e+13 & 1.02e+2 \\  
 -13.41 & 3.55e+14 & 1.22e+13 & 2.44e+14 & 1.07e+13 & 1.09e+14 & 7.17e+12 & 1.11e+2 \\  
 -13.17 & 9.99e+13 & 4.78e+12 & 6.13e+13 & 3.96e+12 & 3.85e+13 & 3.14e+12 & 1.16e+2 \\  
 -12.93 & 2.26e+13 & 1.70e+12 & 1.20e+13 & 1.31e+12 & 1.05e+13 & 1.23e+12 & 1.19e+2 \\  
 -12.69 & 7.51e+12 & 7.40e+11 & 4.09e+12 & 5.57e+11 & 3.41e+12 & 5.08e+11 & 1.21e+2 \\  
 -12.46 & 1.37e+12 & 2.38e+11 & 6.41e+11 & 1.65e+11 & 7.27e+11 & 1.76e+11 & 1.23e+2 \\  
 -12.22 & 4.27e+11 & 1.00e+11 & 1.89e+11 & 6.68e+10 & 2.36e+11 & 7.47e+10 & 1.24e+2 \\  
 -11.74 & 1.56e+10 & 1.10e+10 & 7.45e+09 & 7.45e+09 & 7.45e+09 & 7.45e+09 & 1.25e+2   
\enddata
\tablenotetext{a}{$dN/dS$ has the unit of Deg$^{-2}~(\flux)^{-1}$.}
\end{deluxetable}

\begin{deluxetable}{cccccccc}
\tabletypesize{\scriptsize}
\tablecolumns{8}
\tablewidth{0pt}
\tablecaption{\swift\ Differential AGN Number Counts in the Hard X-ray Band (2--8~keV) \label{tab:agndiffhard}}
\tablehead{
\colhead{$\log{S}$} & \multicolumn{6}{c}{dN/dS}\tablenotemark{a} & \colhead{Area} \\
\colhead{($\flux$)} & \colhead{\swift-All} & \colhead{uncertainty} & \colhead{\swift-\wise\ Type I} & \colhead{uncertainty} & \colhead{\swift-\wise\ Type II} & \colhead{uncertainty} & \colhead{(Deg$^2$)}
}
\startdata
 -14.58 & 2.26e+17 & 2.61e+16 & 1.38e+17 & 3.07e+16 & 8.66e+16 & 2.27e+16 & 3.020e-1 \\ 
 -14.34 & 5.37e+16 & 2.40e+15 & 3.04e+16 & 2.60e+15 & 2.35e+16 & 2.29e+15 & 4.669e+0 \\ 
 -14.10 & 2.03e+16 & 4.97e+14 & 1.29e+16 & 5.22e+14 & 7.13e+15 & 3.94e+14 & 2.218e+1 \\ 
 -13.86 & 8.05e+15 & 1.51e+14 & 5.51e+15 & 1.55e+14 & 2.44e+15 & 1.04e+14 & 5.119e+1 \\ 
 -13.62 & 2.53e+15 & 5.21e+13 & 1.88e+15 & 5.20e+13 & 6.26e+14 & 3.02e+13 & 7.481e+1 \\ 
 -13.38 & 6.29e+14 & 1.74e+13 & 4.95e+14 & 1.69e+13 & 1.30e+14 & 8.76e+12 & 9.322e+1 \\ 
 -13.14 & 1.53e+14 & 6.12e+12 & 1.25e+14 & 5.96e+12 & 2.79e+13 & 2.82e+12 & 1.047e+2 \\ 
 -12.91 & 3.83e+13 & 2.23e+12 & 3.13e+13 & 2.15e+12 & 6.87e+12 & 1.01e+12 & 1.123e+2 \\ 
 -12.67 & 8.32e+12 & 7.69e+11 & 6.18e+12 & 7.09e+11 & 2.12e+12 & 4.15e+11 & 1.174e+2 \\ 
 -12.43 & 1.99e+12 & 2.82e+11 & 1.56e+12 & 2.71e+11 & 4.25e+11 & 1.41e+11 & 1.204e+2 \\ 
 -12.19 & 7.28e+11 & 1.28e+11 & 4.28e+11 & 1.07e+11 & 2.95e+11 & 8.90e+10 & 1.220e+2 \\ 
 -11.95 & 1.03e+11 & 3.67e+10 & 5.13e+10 & 2.56e+10 & 5.13e+10 & 2.56e+10 & 1.231e+2 \\ 
 -11.71 & 2.96e+10 & 1.48e+10 & 2.16e+10 & 1.25e+10 & 7.22e+09 & 7.22e+09 & 1.242e+2 \\ 
 -11.23 & 7.32e+09 & 4.22e+09 & 4.72e+09 & 3.34e+09 & 2.36e+09 & 2.36e+09 & 1.252e+2
\enddata    
\tablenotetext{a}{$dN/dS$ has the unit of Deg$^{-2}~(\flux)^{-1}$.}
\end{deluxetable}

\begin{deluxetable}{ccc}
\tabletypesize{\scriptsize}
\tablecolumns{3}
\tablewidth{0pt}
\tablecaption{\swift\ Cumulative Cluster Number Counts in the 0.5--2~keV Band\label{tab:clcumul}}
\tablehead{
\colhead{$\log{S}$} & \colhead{$N(>S)$} & \colhead{Area} \\
\colhead{($\flux$)} & \colhead{(Deg$^{-2}$)} & \colhead{(Deg$^{2}$)}
}
\startdata
 -14.36 & 21.030  $\pm$   1.494 &  11.36 \\ 
 -14.12 & 16.904  $\pm$   1.112 &  23.21 \\ 
 -13.88 &  9.701  $\pm$   0.625 &  39.80 \\ 
 -13.65 &  4.562  $\pm$   0.339 &  56.07 \\ 
 -13.41 &  2.037  $\pm$   0.184 &  71.09 \\ 
 -13.17 &  1.071  $\pm$   0.122 &  84.49 \\ 
 -12.93 &  0.504  $\pm$   0.075 &  94.75 \\ 
 -12.69 &  0.258  $\pm$   0.049 &  103.0 \\ 
 -12.45 &  0.171  $\pm$   0.039 &  109.4 \\ 
 -12.21 &  0.076  $\pm$   0.025 &  114.4 \\ 
 -11.97 &  0.050  $\pm$   0.020 &  118.1 \\ 
 -11.73 &  0.041  $\pm$   0.018 &  120.4 \\ 
 -11.49 &  0.025  $\pm$   0.014 &  122.2
\enddata
\end{deluxetable}

\begin{deluxetable}{cccccc}
\tabletypesize{\scriptsize}
\tablecolumns{6}
\tablewidth{0pt}
\tablecaption{Fitting Results for \swift\ Differential AGN Number Counts\label{tab:agnpara}}
\tablehead{
\colhead{Data Set} & \colhead{Band} & \colhead{Normalization} & \colhead{Break} & \colhead{a} & \colhead{b} \\
\colhead{} & \colhead{} & \colhead{[Deg$^{-2} (\flux)^{-1}$]} & \colhead{($10^{-15}~\flux$)} & \colhead{} & \colhead{}
}
\startdata
\swift-All             & 0.5--8~keV &  531.91$\pm$  250.04 &    3.67$\pm$ 1.57 &    1.34\tablenotemark{a} &    2.37$\pm$    0.01 \\
\swift-\wise\ MIR-red  & 0.5--8~keV &  185.08$\pm$   13.09 &    8.10\tablenotemark{a} &    1.34\tablenotemark{a} &    2.51$\pm$    0.02 \\
\swift-\wise\ MIR-blue & 0.5--8~keV &  135.34$\pm$   10.39 &    8.10\tablenotemark{a} &    1.34\tablenotemark{a} &    2.39$\pm$    0.02 \\
\swift-All             & 0.5--2~keV & 108.60$\pm$    9.78 &   10.08$\pm$    1.40 &    1.49\tablenotemark{a} &    2.54$\pm$    0.04 \\
\swift-\wise\ MIR-red  & 0.5--2~keV &  66.15$\pm$   11.80 &    6.00\tablenotemark{a} &    1.49\tablenotemark{a} &    2.46$\pm$    0.06 \\
\swift-\wise\ MIR-blue & 0.5--2~keV &  47.81$\pm$    4.73 &    6.00\tablenotemark{a} &    1.49\tablenotemark{a} &    2.42$\pm$    0.04 \\
\swift-All             & 2--8~keV   & 578.61\tablenotemark{b}              &    2.15\tablenotemark{b}              &    1.32\tablenotemark{a} &    2.39$\pm$    0.14 \\
\swift-\wise\ MIR-red  & 2--8~keV   & 168.06$\pm$   26.30 &    6.40\tablenotemark{a} &    1.32\tablenotemark{a} &    2.46$\pm$    0.04 \\
\swift-\wise\ MIR-blue & 2--8~keV   &  63.95$\pm$   12.15 &    6.40\tablenotemark{a} &    1.32\tablenotemark{a} &    2.50$\pm$    0.06 
\enddata
\tablenotetext{a}{This parameter is fixed the same as the CDF-S measurement.}
\tablenotetext{b}{The parameters are unconstrained.}
\tablecomments{The data are fit by a broken powerlaw with a sharp break as Equation~\ref{eq:agn}.}
\end{deluxetable}

\begin{deluxetable}{ccccc}
\tabletypesize{\scriptsize}
\tablecolumns{5}
\tablewidth{0pt}
\tablecaption{Fitting Results for Cumulative Cluster Number Counts in the 0.5--2~keV Band\label{tab:clpara}}
\tablehead{
\colhead{Data Set} & \colhead{Normalization} & \colhead{Break} & \colhead{a} & \colhead{b} \\
\colhead{} & \colhead{(Deg$^{-2}$)} & \colhead{($10^{-15}~\flux$)} & \colhead{} & \colhead{}
}
\startdata
\swift &   27.73$\pm$   2.04 &   5.10\tablenotemark{a} &    0.52\tablenotemark{a} &    1.24$\pm$    0.04 \\
All    &   28.13$\pm$   0.55 &   5.10$\pm$    1.64     &    0.52$\pm$    0.06     &    1.22$\pm$    0.01 
\enddata
\tablenotetext{a}{This parameter is fixed as the fitting result using all the data.}
\tablecomments{The data are fit by a smooth broken powerlaw as Equation~\ref{eq:cl} with the smoothness parameter fixed as $c=2$.}
\end{deluxetable}

\end{document}